\documentclass[aps,pre,showpacs,amsmath,amssymb]{revtex4-1}

\usepackage{graphicx}

\begin{document}

\title{Tsallis distributions and $1/f$ noise from nonlinear stochastic
differential equations}

\author{J.~Ruseckas}

\email{julius.ruseckas@tfai.vu.lt}

\homepage{http://www.itpa.lt/~ruseckas}

\author{B.~Kaulakys}

\affiliation{Institute of Theoretical Physics and Astronomy, Vilnius University,
A.~Go\v{s}tauto 12, LT-01108 Vilnius, Lithuania}

\begin{abstract} 
Probability distributions which emerge from the formalism of nonextensive
statistical mechanics have been applied to a variety of problems. In this paper
we unite modeling of such distributions with the model of widespread $1/f$
noise. We propose a class of nonlinear stochastic differential equations giving
both the $q$-exponential or $q$-Gaussian distributions of signal intensity,
revealing long-range correlations and $1/f^{\beta}$ behavior of the power
spectral density. The superstatistical framework to get $1/f^{\beta}$ noise with
$q$-exponential and $q$-Gaussian distributions of the signal intensity in is
proposed, as well.
\end{abstract}


\maketitle

\section{Introduction}

Stationary stochastic processes and signals are prevalent across many fields of
science and engineering. Many complex systems show large fluctuations of
macroscopic quantities that follow non-Gaussian, heavy-tailed, power-law
distributions with the power-law temporal correlations, scaling, and the
(multi)fractal features \cite{Mandelbrot-1999,Mantegna-2001,Lowen-2005}. The
power-law distributions, scaling, self-similarity and fractality are sometimes
related both with the nonextensive statistical mechanics
\cite{tsallis-1988,queiros-2005-3,Tsallis-2009,tsallis-2009a,telesca-2010} and
with the power-law behavior of the power spectral density, i.e., $1/f^{\beta}$
noise (see, e.g., \cite{Lowen-2005,kaulakys-2005,kaulakys-2009,Fossion-2010},
and references herein).

There exist a number of systems, involving long-range interactions, long-range
memory and anomalous diffusion, that possess anomalous properties in view of
traditional Boltzmann-Gibbs statistical mechanics. Nonextensive statistical
mechanics represents a consistent theoretical background for the investigation
of some properties, like fractality, multifractality, self-similarity,
long-range dependencies and so on, of such complex systems,
\cite{Tsallis-2009,tsallis-2009a,telesca-2010}. Concepts related with
nonextensive statistical mechanics have found applications in a variety of
disciplines including physics, chemistry, biology, mathematics, economics,
informatics, and the interdisciplinary field of complex systems (see, e.g.,
\cite{gell-mann-2004,abe-2006,Picoli-2009} and references herein).

The nonextensive statistical mechanics framework is based on the entropic form
\cite{tsallis-1988,Tsallis-2009}
\begin{equation}
S_q=\frac{1-\int_{-\infty}^{+\infty}[p(z)]^qdz}{q-1}\,, 
\label{eq:q-entr}
\end{equation} 
where $p(z)$ is the probability density of finding the system with the parameter
$z$. Entropy (\ref{eq:q-entr}) is an extension of the Boltzmann-Gibbs entropy
$S_{\mathrm{BG}}=-\int_{-\infty}^{+\infty}p(z)\ln p(z)dz$, which restores from
Eq.~(\ref{eq:q-entr}) at $q=1$ \cite{Tsallis-2009,tsallis-2009a}. 

By applying the standard variational principle on entropy (\ref{eq:q-entr}) with
the constraints $\int_{-\infty}^{+\infty}p(z)dz=1$ and 
\begin{equation}
\frac{\int_{-\infty}^{+\infty}z^2[p(z)]^qdz}{\int_{-\infty}^{
+\infty}[p(z)]^qdz}=\sigma_q^2\,,
\end{equation}
where $\sigma_{q}^2$ is the generalized second-order moment
\cite{tsallis-1998,prato-1999,tsallis-1999}, one obtains the $q$-Gaussian
distribution density
\begin{equation}
p(z)=A\exp_q(-Bz^2)\,.
\end{equation}
Here $\exp_q(\cdot)$ is the $q$-exponential function defined as 
\begin{equation}
\exp_q(x)\equiv [1+(1-q)x]_{+}^{\frac{1}{1-q}}\,, 
\label{eq:q-exp1}
\end{equation}
with $[(\ldots)]_{+}=(\ldots)$ if $(\ldots)>0$, and zero otherwise.
Asymptotically, as $x\rightarrow \infty$, $\exp_q(x)\sim x^{-\lambda}$, where
$\lambda=(q-1)^{-1}$, i.e., we have the power-law distribution. The (more)
generalized entropies and distribution functions are introduced in
Refs.~\cite{hanel-2011a,hanel-2011b}. 

Statistics associated to Eqs.~(\ref{eq:q-entr})--(\ref{eq:q-exp1}) has been
successfully applied to phenomena with the scale-invariant geometry, like in
low-dimensional dissipative and conservative maps
\cite{lyra-1998,baldovin-2000}, anomalous diffusion \cite{borland-1998},
turbulent flows \cite{beck-2001}, Langevin dynamics with fluctuating temperature
\cite{beck-2003,wilk-2000,beck-2001-1}, long-range many-body classical
Hamiltonians \cite{latora-2001}, and to the financial systems
\cite{tsallis-2003a,drozdz-2010}.

For the modeling of distributions of the nonextensive statistical mechanics, the
nonlinear Fokker-Planck equations and corresponding nonlinear stochastic
differential equations (SDEs) \cite{borland-1998,borland-2002}, SDEs with
additive and multiplicative noises \cite{anteneodo-2003,santos-2010}, with
multiplicative noise only \cite{queiros-2007}, and with fluctuating friction
forces \cite{beck-2001-1} have been proposed. However, the exhibition of the
long-range correlations and $1/f^{\beta}$ noise has not been observed.

The phrases ``$1/f$ noise'', ``$1/f$ fluctuations'', and ``flicker noise'' refer
to the phenomenon, having the power spectral density at low frequencies $f$ of
signals of the form $S(f)\sim 1/f^{\beta}$, with $\beta$ being a
system-dependent parameter. Signals with $0.5<\beta<1.5$ are found widely in
nature, occurring in physics, electronics, astrophysics, geophysics, economics,
biology, psychology, language and even music
\cite{scholarpedia-2009,weissman-1988,Gilden-1995,milotti-2002,wong-2003} (see
also references in paper \cite{kaulakys-2009}). The case of $\beta=1$, or ``pink
noise'', is the one of the most interesting. The widespread occurrence of
processes exhibiting $1/f$ noise suggests that a generic, at least mathematical
explanation of such phenomena might exist.

One common way for describing stochastic evolution and properties of complex
systems, is by means of generalized stochastic differential equations of motion
\cite{Gardiner04,Risken89,Farias09}. These nondeterministic equations of motion
are used in many systems of interest, such as simulating the Brownian motion in
statistical mechanics, in fundamental aspects of synergetics and biological
systems, field theory models, the financial systems, and in other areas
\cite{Gardiner04,Lax06,Mantegna-2001,Jeanblanc09}.

The purpose of this paper is to model together both, the Tsallis distributions
and $1/f$ noise, using the same nonlinear stochastic differential equations. The
superstatistical approach for modeling of such processes is proposed, as well.

We considered a class of nonlinear stochastic differential equations giving the
power-law behavior of the probability density function (PDF) of the signal
intensity and of the power spectral density ($1/f^{\beta}$ noise) in any
desirably wide range of frequency. Modifications these equations by introduction
of an additional parameter yields Brownian-like motion for small values of the
signal and avoids power-law divergence of the signal distribution, while
preserving $1/f^{\beta}$ behavior of the power spectral density. The PDF of the
signal generated by modified SDEs is $q$-exponential or $q$-Gaussian distribution of
the nonextensive statistical mechanics. The superstatistical framework using a
fast dynamics with the slowly changing parameter described by nonlinear
stochastic differential equations can retain $1/f^{\beta}$ behavior of the power
spectral density as well. When the PDF of the rapidly changing variable is
exponential or Gaussian, we obtain $q$-exponential or $q$-Gaussian long-term
stationary PDF of the signal, respectively.

\section{Nonlinear stochastic differential equation generating asymptotically 
power-law signals with $1/f^{\beta}$ noise} 

\label{sec:nonlin-sde}Starting from the point process model, proposed and
analyzed in
Refs.~\cite{kaulakys-1998,kaulakys-1999,kaulakys-2001,kaulakys-2002,kaulakys-2003,kaulakys-2005},
the nonlinear SDEs generating processes with $1/f^{\beta}$ noise are derived
\cite{kaulakys-2004,kaulakys-2006,kaulakys-2009}. The general expression for the
proposed class of It\^o SDEs is
\begin{equation}
dx=\sigma^2\left(\eta-\frac{1}{2}\lambda\right)x^{2\eta-1}dt+\sigma
x^{\eta}dW\,.
\label{eq:sde}
\end{equation}
Here $x$ is the signal, $\eta\neq1$ is the power-law exponent of the
multiplicative noise, $\lambda$ defines the behavior of stationary probability
distribution, and $W$ is a standard Wiener process (the Brownian motion).

The nonlinear SDE (\ref{eq:sde}) has the simplest form of the multiplicative
noise term, $\sigma x^{\eta}dW$. Equations with multiplicative noise and with
the drift coefficient proportional to the Stratonovich drift correction for
transformation from the Stratonovich to the It\^o stochastic equation
\cite{arnold-2000}, generate signals with the power-law distributions
\cite{kaulakys-2009}. Equation (\ref{eq:sde}) is of such a type. Therefore, the
relationship between the exponents in the drift term, $2\eta-1$, and in the
noise term, $\eta$, of these equations follows from the requirement of modeling
the signals with the power-law distributions. More reasoning of the correlation
of these exponents, and of the type of equations like (\ref{eq:sde}) in general,
have been given in
Refs.~\cite{gontis-2004,kaulakys-2004,kaulakys-2005,kaulakys-2006,kaulakys-2009,kaulakys-2009-3}. 

On the other hand, the simple transformation of the variable, $y=x^{\alpha}$, 
gives equation of the same type (\ref{eq:sde}) only with different parameters,
$\sigma^{\prime} = \alpha\sigma$, $\eta^{\prime} = (\eta-1)/\alpha + 1$, 
and $\lambda^{\prime} = (\lambda-1)/\alpha + 1$.
E.g., for $\alpha = 1-\eta $ with $\eta\neq1$ we get $\eta^{\prime} = 0$, i.e.,
equation for the variable $y$ having additive noise and nonlinear drift, the
well-known in econophysics and finance SDE describing the Bessel process
\cite{Jeanblanc09}. Thus the observable $x$ may be a function of
another variable $y$, described by a simpler SDE with additive noise.

Nonlinear SDE, corresponding to a particular case of (\ref{eq:sde}) with
$\eta=0$, i.e., with linear noise and non-linear drift, was considered in
Ref.~\cite{mamontov-1997}. It has been found that if the damping decrease with
increasing $|x|$, then the solution of such a nonlinear SDE has long correlation
time. The connection of the power spectral density of the signal generated by
SDE (\ref{eq:sde}) with the behavior of the eigenvalues of the corresponding
Fokker-Planck equation was analyzed in Ref.~\cite{ruseckas-2010}. This
connection was generalized in Ref.~\cite{Erland2011} where it has been shown
that $1/f^{\beta}$ noise is equivalent to a Markovian eigenstructure power
relation.

In order to obtain a stationary process and avoid the divergence of steady state
PDF the diffusion of stochastic variable $x$ should be restricted at least from
the side of small values, or equation (\ref{eq:sde}) should be modified. The
Fokker-Planck equation corresponding to SDE (\ref{eq:sde}) with restrictions of
diffusion of stochastic variable $x$ gives the power-law steady state PDF
\begin{equation}
P(x)\sim x^{-\lambda}
\end{equation}
with the exponent $\lambda$, when the variable $x$ is far from the ends of the
diffusion interval. The simplest choice of the restriction is the reflective
boundary conditions at $x=x_{\mathrm{min}}$ and $x=x_{\mathrm{max}}$.
Exponentially restricted diffusion with the steady state PDF
\begin{equation}
P(x)\sim\frac{1}{x^{\lambda}}\exp\left\{
-\left(\frac{x_{\mathrm{min}}}{x}\right)^m-\left(\frac{x}{x_{\mathrm{
max}}}\right)^m\right\}
\end{equation}
is generated by the SDE
\begin{equation}
dx=\sigma^2\left[\eta-\frac{1}{2}\lambda+\frac{m}{2}\left(\frac{x_{\mathrm{
min}}^m}{x^m}-\frac{x^m}{x_{\mathrm{max}}^m}\right)\right]x^{2\eta-1}dt+\sigma
 x^{\eta}dW
\label{eq:sde-restricted}
\end{equation}
obtained from Eq.~(\ref{eq:sde}) by introducing the additional terms.

In Refs.~\cite{kaulakys-2005,kaulakys-2006} it was
shown that SDE (\ref{eq:sde}) generates signals with power spectral density
\begin{equation}
S(f)\sim
\frac{1}{f^{\beta}}\,,\qquad\beta=1+\frac{\lambda-3}{2(\eta-1)}\,.
\label{eq:beta}
\end{equation}
in a wide interval of frequencies. SDE (\ref{eq:sde}) exhibits the following
scaling property: changing the stochastic variable from $x$ to a scaled variable
$x^{\prime}=ax$ changes the time-scale of the equation to
$t^{\prime}=a^{2(1-\eta)}t$ leaving the form of the equation unchanged. This
scaling property is one of the reasons for the appearance of the $1/f^{\beta}$
power spectral density.

\begin{figure}
\includegraphics[width=0.45\textwidth]{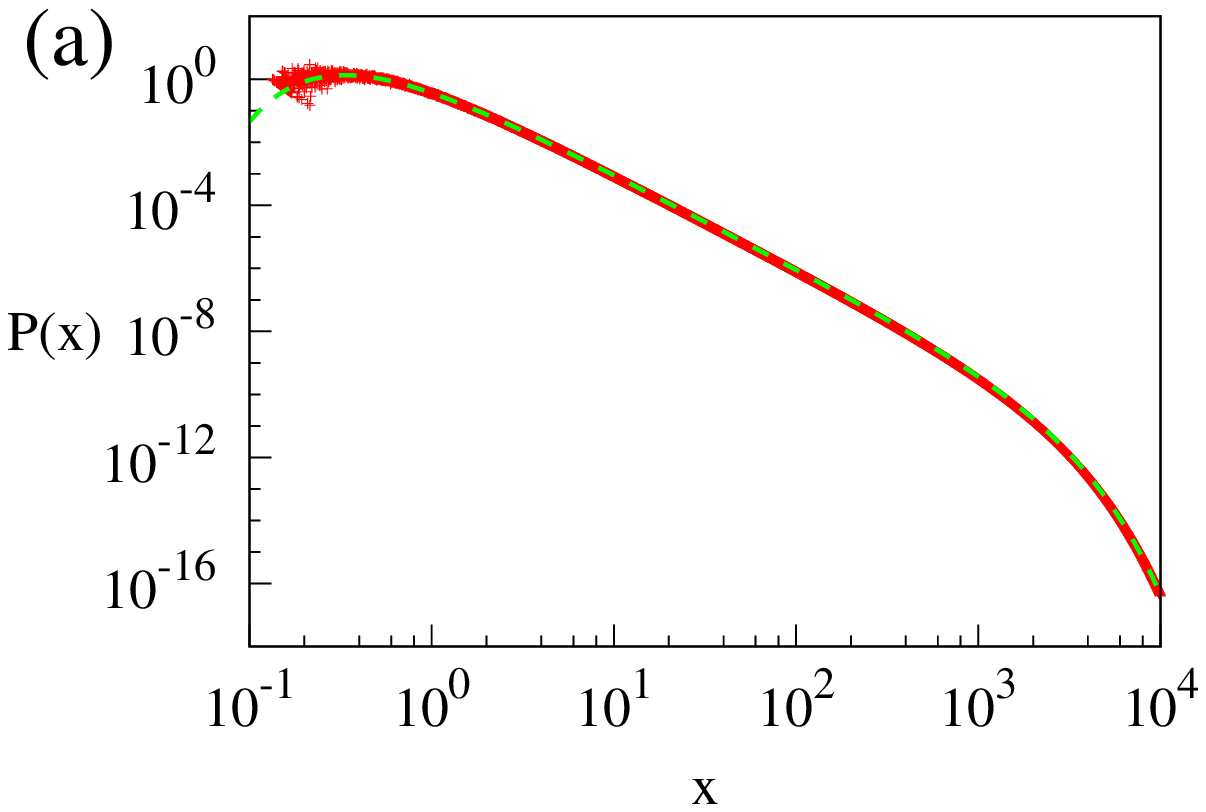}
\includegraphics[width=0.45\textwidth]{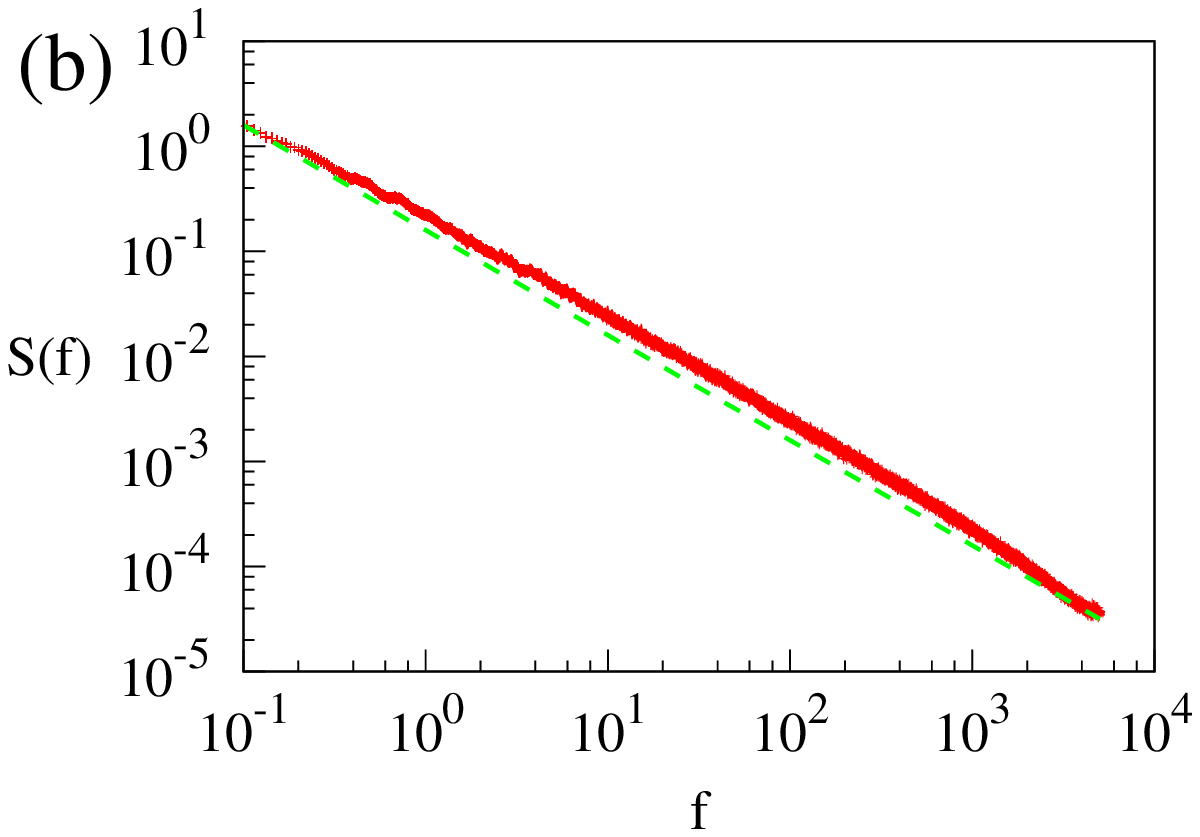}
\caption{(Color online) (a) Steady state PDF $P(x)$ of the signal generated by
Eq.~(\ref{eq:sde1}) with $x_{\mathrm{min}}=1$ and $x_{\mathrm{max}}=1000$. The
dashed (green) line is the analytical expression for the steady state PDF. (b)
Power spectral density $S(f)$ of the same signal. The dashed (green) line shows
the slope $1/f$.}
\label{fig:sde}
\end{figure}

For $\lambda=3$ we get that $\beta=1$ and SDE (\ref{eq:sde}) should give signal
exhibiting $1/f$ noise. One example of the equation (\ref{eq:sde-restricted})
with $\lambda=3$, $m=1$, $\sigma=1$, and $\eta=5/2$ is
\begin{equation}
dx=\left[1+\frac{1}{2}\left(\frac{x_{\mathrm{min}}}{x}-\frac{x}{x_{\mathrm{
max}}}\right)\right]x^4dt+x^{5/2}dW\,.
\label{eq:sde1}
\end{equation}
Note, that $\eta=5/2$ corresponds to the simplest point process model
with the Brownian motion of the interevent time $\tau_k$ in the
events space ($k$-space) \cite{kaulakys-2004,kaulakys-2005,kaulakys-2006} , 
\begin{equation}
d\tau_k= \sigma dW_k\,. 
\label{eq:point}
\end{equation}
Consequently, the simple point process model may provide one possible reasoning
of use of the strongly nonlinear multiplicative SDEs for modeling of long-range
correlated systems.

Comparison of numerically obtained steady state PDF and power spectral density
with analytical expressions is presented in Fig.~\ref{fig:sde}. For the
numerical solution we use Euler-Marujama approximation with variable step of
integration, transforming the differential equations to the difference equations
\cite{kaulakys-2004,kaulakys-2006}. We see a good agreement of the numerical
results with the analytical expressions. Numerical solution of the equations
confirms the presence of the frequency region for which the power spectral
density has $1/f^{\beta}$ dependence. The width of this region can be increased
by increasing the ratio between the minimum and the maximum values of the
stochastic variable $x$. In addition, the region in the power spectral density
with the power-law behavior depends on the exponent $\eta$: if $\eta=1$ then
this width is zero; the width increases with increasing the difference
$|\eta-1|$ \cite{ruseckas-2010}.

The numerical analysis of the proposed SDE~(\ref{eq:sde}) reveals the secondary
structure of the signal composed of peaks or bursts, corresponding to the large
deviations of the variable $x$ from the proper average fluctuations
\cite{kaulakys-2009}. Bursts are characterized by power-law distributions of
burst size, burst duration, and interburst time.

\section{Stochastic differential equations giving $q$-distributions}

The power spectral density of the form $1/f^{\beta}$ is determined mainly by
power-law behavior of the coefficients of SDEs~(\ref{eq:sde}),
(\ref{eq:sde-restricted}) at large values of $x\gg x_{\mathrm{min}}$. Changing
the coefficients at small $x$, the spectrum preserves the power-law behavior. In
addition, Fokker-Planck equation corresponding to SDE~(\ref{eq:sde}) gives the 
steady state PDF with power-law dependence on $x$ as does the $q$-exponential
function for large $x$. Therefore, SDE~(\ref{eq:sde}) can be modified to yield
generalized canonical distributions of nonextensive statistical mechanics.

\subsection{$q$-exponential distribution}

Modified stochastic differential equation
\begin{equation}
dx=\sigma^2\left(\eta-\frac{1}{2}\lambda\right)(x+x_0)^{2\eta-1}dt+\sigma(x
+x_0)^{\eta}dW
\label{eq:sde-q-exp}
\end{equation}
with the reflective boundary condition at $x=0$ was considered in
\cite{kaulakys-2009}. The Fokker-Planck equation corresponding to SDE
(\ref{eq:sde-q-exp}) for $x\geq 0$ gives $q$-exponential steady state PDF
\begin{equation}
P(x)=\frac{\lambda-1}{x_0}\left(\frac{x_0}{x+x_0}\right)^{\lambda}=\frac{
\lambda-1}{x_0}\exp_q(-\lambda x/x_0)\,,\qquad
q=1+1/\lambda\,.
\label{eq:pdf-q-exp}
\end{equation}
The addition of parameter $x_0$ restricts the divergence of the power-law
distribution of $x$ at $x\rightarrow0$. Equation (\ref{eq:sde-q-exp}) for
small $x\ll x_0$ represents the linear additive stochastic process generating
the Brownian motion with the steady drift, while for $x\gg x_0$ it reduces to
the multiplicative SDE~(\ref{eq:sde}). This modification of the SDE retains the
frequency region with $1/f^{\beta}$ behavior of the power spectral density.

\begin{figure}
\includegraphics[width=0.4\textwidth]{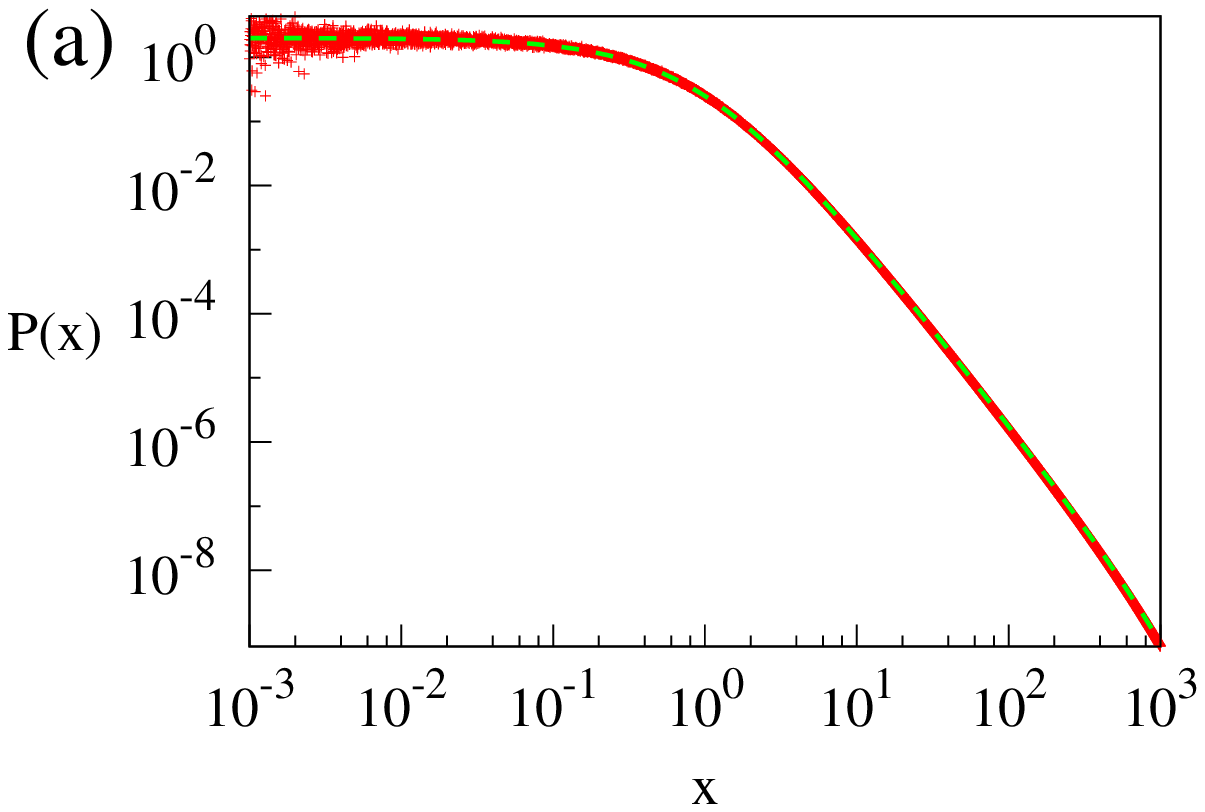}\includegraphics[width=0.4\textwidth]{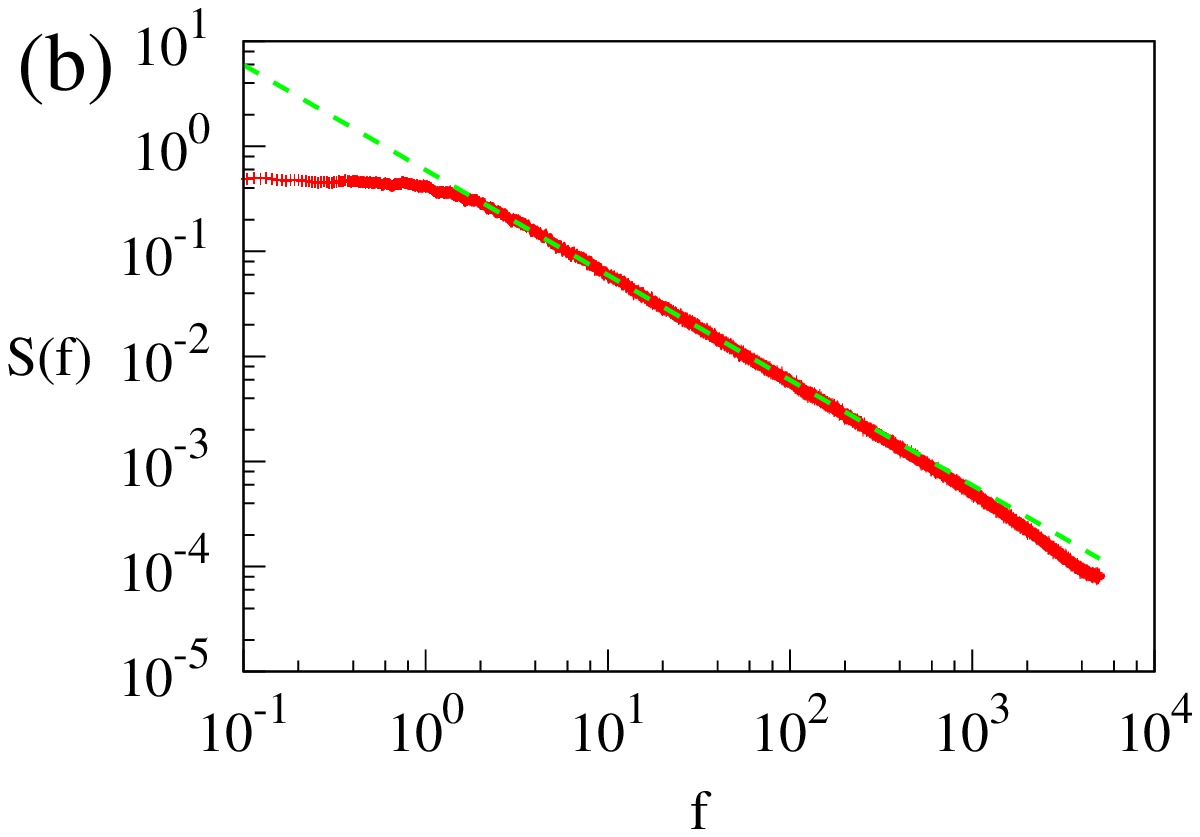}
\caption{(Color online) (a) Steady state PDF $P(x)$ of the signal generated by
Eq.~(\ref{eq:sde-q-exp}). The dashed (green) line is the analytical
$q$-exponential expression (\ref{eq:pdf-q-exp}) for the steady state PDF. (b)
Power spectral density $S(f)$ of the same signal. The dashed (green) line shows
the slope $1/f$. The parameters used are $\lambda=3$, $\eta=5/2$, $x_0=1$, and
$\sigma=1$.}
\label{fig:sde-q-exp}
\end{figure}

Comparison of numerically obtained steady state PDF and power spectral density
with analytical expressions is presented in Fig.~\ref{fig:sde-q-exp}. We see a
good agreement of the numerical results with the analytical expressions.
Numerical solution confirms the presence of the frequency region where the
power spectral density has $1/f^{\beta}$ dependence. The lower bound of this
frequency region depends on the parameter $x_0$.

\subsection{$q$-Gaussian distribution}

\label{sub:q-gauss}Stochastic differential equation
\begin{equation}
dx=\sigma^2\left(\eta-\frac{1}{2}\lambda\right)(x^2+x_0^2)^{\eta-1}xdt
+\sigma(x^2+x_0^2)^{\eta/2}dW
\label{eq:sde-q-gauss}
\end{equation}
in contrast to all other equations analyzed in this article, allows negative
values of $x$. This equation was introduced in
Refs.~\cite{kaulakys-2009-2,gontis-2009,gontis-2010}. The simple case $\eta=1$
is used in the model of return in Ref.~\cite{queiros-2007}. Note, that $\eta=1$
does not give $1/f^{\beta}$ power spectral density. The Fokker-Planck equation
corresponding to SDE~(\ref{eq:sde-q-gauss}) gives $q$-Gaussian steady state PDF
\begin{eqnarray}
P(x) & = &
\frac{\Gamma\left(\frac{\lambda}{2}\right)}{\sqrt{\pi}x_0\Gamma\left(\frac{
\lambda-1}{2}\right)}\left(\frac{x_0^2}{x_0^2+x^2}\right)^{\frac{\lambda}{
2}}=\frac{\Gamma\left(\frac{\lambda}{2}\right)}{\sqrt{\pi}x_0\Gamma\left(\frac{
\lambda-1}{2}\right)}\exp_q\left(-\lambda\frac{x^2}{2x_0^2}\right)\,,
\label{eq:pdf-q-gauss}
\\ q & = & 1+2/\lambda\,.\nonumber
\end{eqnarray}
The addition of parameter $x_0$ restricts the divergence of the power-law
distribution of $x$ at $x\rightarrow0$. Equation (\ref{eq:sde-q-gauss}) for
small $|x|\ll x_0$ represents the linear additive stochastic process generating
the Brownian motion with the linear relaxation, while for $x\gg x_0$ it reduces
to the multiplicative SDE~(\ref{eq:sde}). This modification of the SDE, even the
introduction of negative values of the stochastic variable $x$, does not destroy
the frequency region with $1/f^{\beta}$ behavior of the power spectral density.

\begin{figure}
\includegraphics[width=0.4\textwidth]{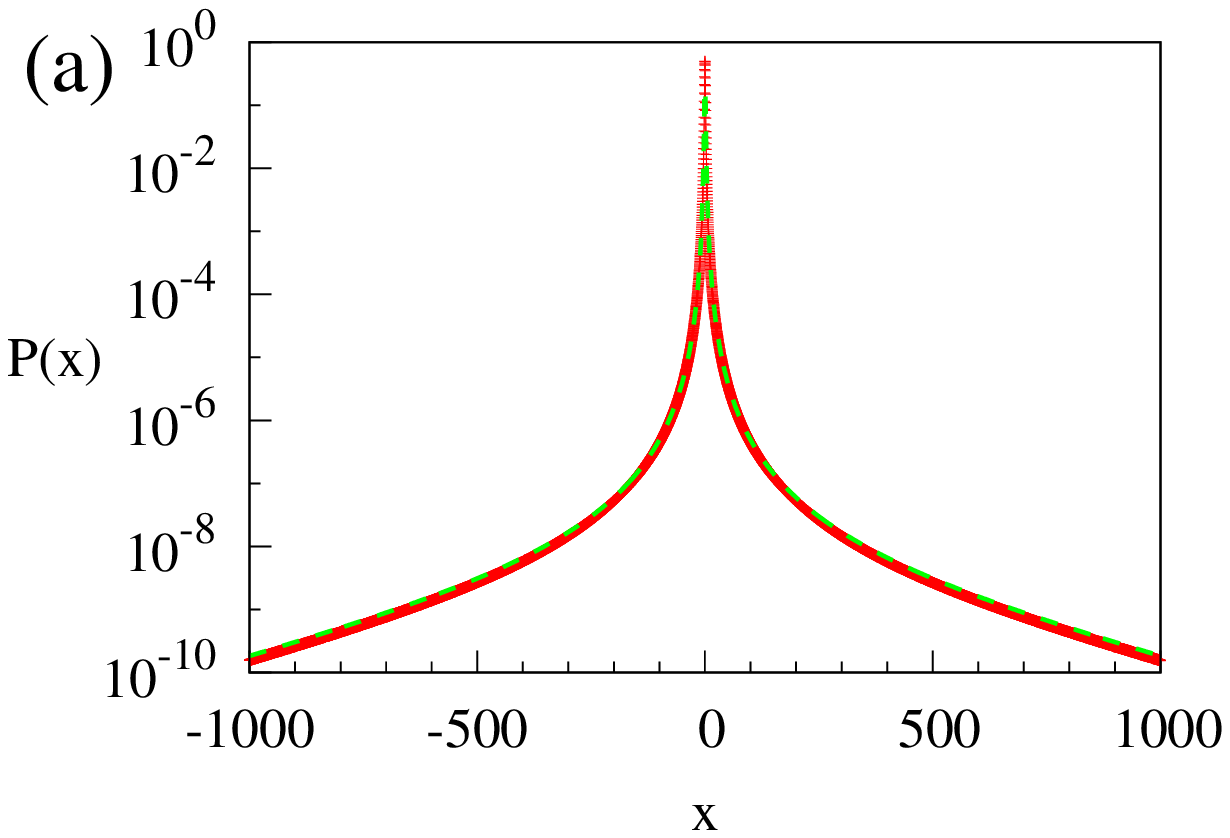}\includegraphics[width=0.4\textwidth]{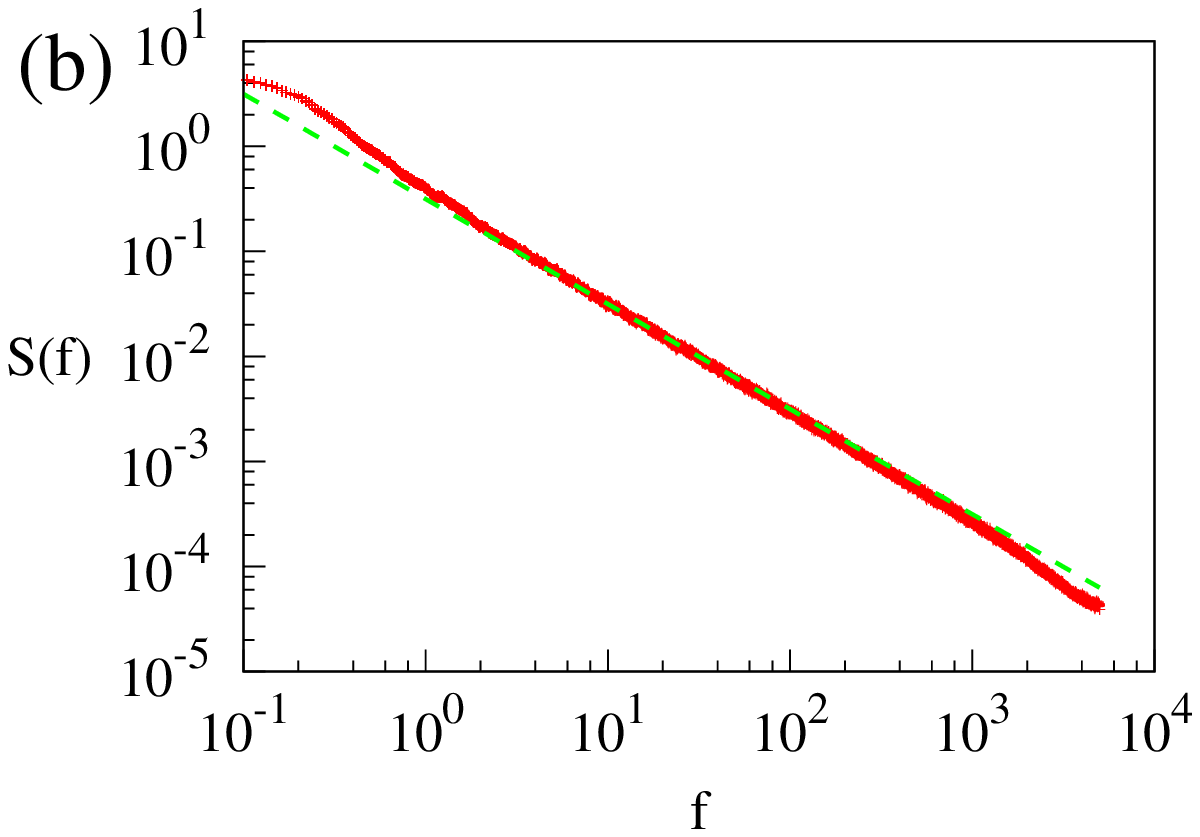}
\caption{(Color online) (a) Steady state PDF $P(x)$ of the signal generated by
Eq.~(\ref{eq:sde-q-gauss}). The dashed (green) line is the analytical
$q$-Gaussian expression (\ref{eq:pdf-q-gauss}) for the steady state PDF. (b)
Power spectral density $S(f)$ of the same signal. The dashed (green) line shows
the slope $1/f$. The parameters used are $\lambda=3$, $\eta=5/2$, $x_0=1$, and
$\sigma=1$.}
\label{fig:sde-q-gauss}
\end{figure}

Comparison of numerically obtained steady state PDF and power spectral density
with analytical expressions is presented in Fig.~\ref{fig:sde-q-gauss}. A good
agreement of the numerical results with the analytical expressions is found.
Numerical solution confirms the presence of the frequency region where the
power spectral density has $1/f^{\beta}$ dependence.

\section{Superstatistics and $1/f^{\beta}$ noise}

Many non-equilibrium systems exhibit spatial or temporal fluctuations of some
parameter. There are two time scales: the scale in which the dynamics is able to
reach a stationary state, and the scale at which the fluctuating parameter
evolves. A particular case is when the time needed for the system to reach
stationarity is much smaller than the scale at which the fluctuating parameter
changes. In the long-term, the non-equilibrium system is described by the
superposition of different local dynamics at different time intervals, that has
been called superstatistics
\cite{beck-2003,tsallis-2003,abe-2007,hahn-2010,beck-2011}. The superstatistical
framework has successfully been applied on a widespread of problems like:
interactions between hadrons from cosmic rays \cite{wilk-2000}, fluid turbulence
\cite{beck-2001-1,beck-2005,beck-2005-1,beck-2007}, granular material
\cite{beck-2006}, electronics \cite{sattin-2002}, and economics
\cite{ausloos-2003,queiros-2005,queiros-2004,queiros-2005-1,queiros-2005-2,souza-2006}.

In this article we will consider the case when the fluctuating parameter
$\bar{x}$ evolves according to earlier introduced SDE (\ref{eq:sde-restricted}).
The parameter $\bar{x}$ changes slowly and can by taken as a constant through a
period of time $T$. Due to the scaling properties of Eq.~(\ref{eq:sde}),
mentioned in Sec.~\ref{sec:nonlin-sde}, the characteristic time scale in
Eq.~(\ref{eq:sde}) decreases as a power of $x$. In order to avoid short time
scales and rapid changes of the parameter $\bar{x}$, the possible values of
$\bar{x}$ should be restricted from above. If the maximum value of the parameter
$\bar{x}$ is $\bar{x}_{\mathrm{max}}$, then the time $T$ during which the
parameter $\bar{x}$ changes slowly decreases with increase of
$\bar{x}_{\mathrm{max}}$. Within time scale $T$ the signal $x$ has local
stationary PDF $\varphi(x|\bar{x})$. The long-term stationary PDF of the signal
$x$ is determined as
\begin{equation}
P(x)=\int_0^{\infty}\varphi(x|\bar{x})p(\bar{x})d\bar{x}\,.
\label{eq:super}
\end{equation}
We can expect that at small frequencies $\omega\ll T^{-1}$ the spectrum of the
signal $x$ is determined mainly by the driving SDE. Therefore, we can get the
distribution $P(x)$ determined by Eq.~(\ref{eq:super}) and $1/f^{\beta}$ power
spectral density in a wide region of frequencies. Using the superstatistical
approach, from SDE~(\ref{eq:sde-restricted}) with the exponential restriction of
diffusion, it is possible to obtain the Tsallis probability distributions. 

\subsection{$q$-exponential distribution}

In order to obtain $q$-exponential long-term PDF of the signal $x$ we will
consider the local stationary PDF conditioned to value of the parameter
$\bar{x}$ in the form of exponential distribution
\begin{equation}
\varphi(x|\bar{x})=\bar{x}^{-1}\exp(-x/\bar{x})\,.
\label{eq:exp}
\end{equation}
A Poissonian-like process with slowly diffusing time-dependent average
interevent time was considered in Ref.~\cite{kaulakys-2009-3}. The mean
$\bar{x}$ of the distribution $\varphi(x|\bar{x})$ obeys SDE with exponential
restriction of diffusion, 
\begin{equation}
d\bar{x}=\sigma^2\left[\eta-\frac{\lambda}{2}+\frac{1}{2}\frac{x_0}{\bar{x}}
-\frac{1}{2}\frac{\bar{x}}{\bar{x}_{\mathrm{max}}}\right]\bar{x}^{2\eta-1}dt
+\sigma\bar{x}^{\eta}dW\,.
\label{eq:sde-mean-qexp}
\end{equation}
Here $x_0$ is a parameter describing exponential cut-off of the steady state
PDF of $\bar{x}$ at small values of $\bar{x}$ and the parameter
$\bar{x}_{\mathrm{max}}\gg x_0$ leads to exponential cut-off at large values of
$\bar{x}$. When $\bar{x}\ll\bar{x}_{\mathrm{max}}$ the influence of the
exponential cut-off at large values of $\bar{x}$ is small. Neglecting
$\bar{x}_{\mathrm{max}}$ the steady state PDF from the Fokker-Planck equation
corresponding to Eq.~(\ref{eq:sde-mean-qexp}) is
\begin{equation}
p(\bar{x})=\frac{1}{x_0\Gamma(\lambda-1)}\left(\frac{x_0}{\bar{x}}\right)^{
\lambda}\exp\left(-\frac{x_0}{\bar{x}}\right)\,.
\label{eq:pdf-exp-s}
\end{equation}
Using Eqs.~(\ref{eq:super}), (\ref{eq:exp}), and (\ref{eq:pdf-exp-s}), 
we get that for $x\ll\bar{x}_{\mathrm{max}}$ the 
long-term stationary PDF of signal $x$ is $q$-exponential function, 
\begin{equation}
P(x)=\frac{\lambda-1}{x_0}\left(\frac{x_0}{x+x_0}\right)^{\lambda}=\frac{
\lambda-1}{x_0}\exp_q(-\lambda x/x_0)\,,\qquad
q=1+1/\lambda\,.
\label{eq:pdf-q-exp-s}
\end{equation}

\begin{figure}
\includegraphics[width=0.4\textwidth]{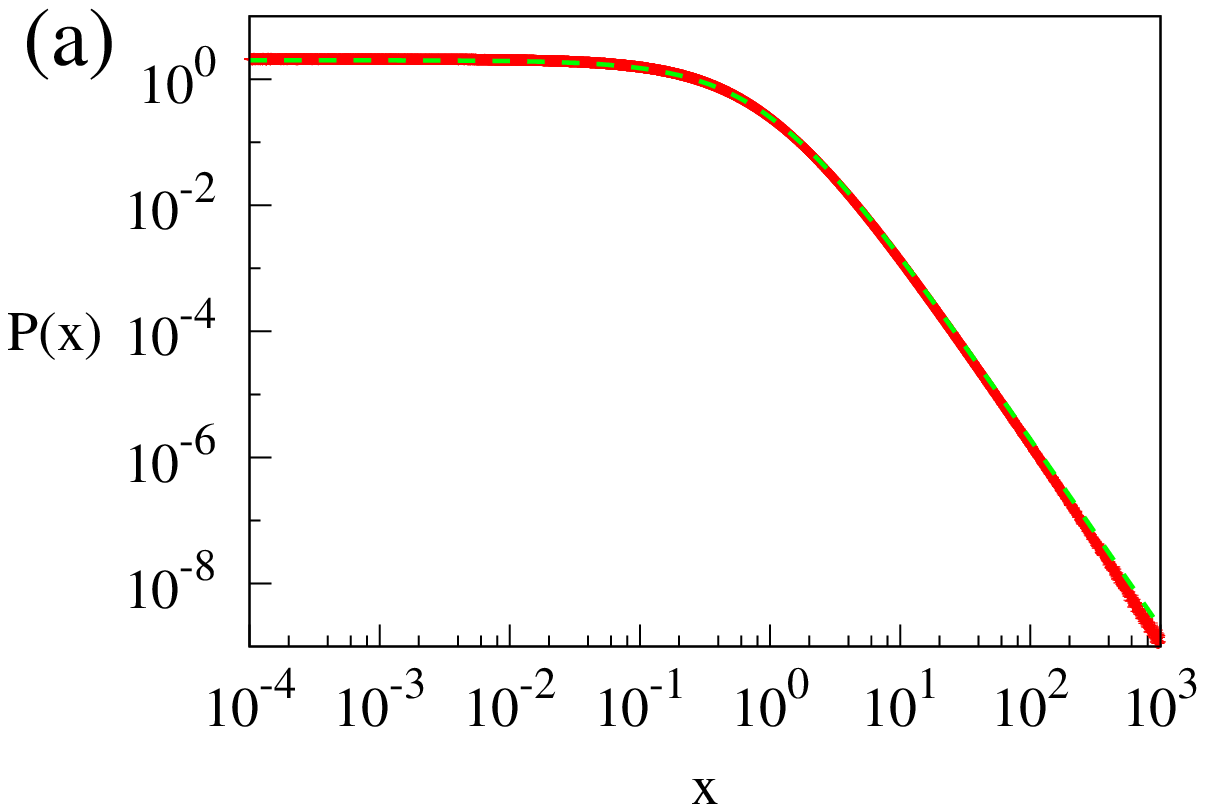}\includegraphics[width=0.4\textwidth]{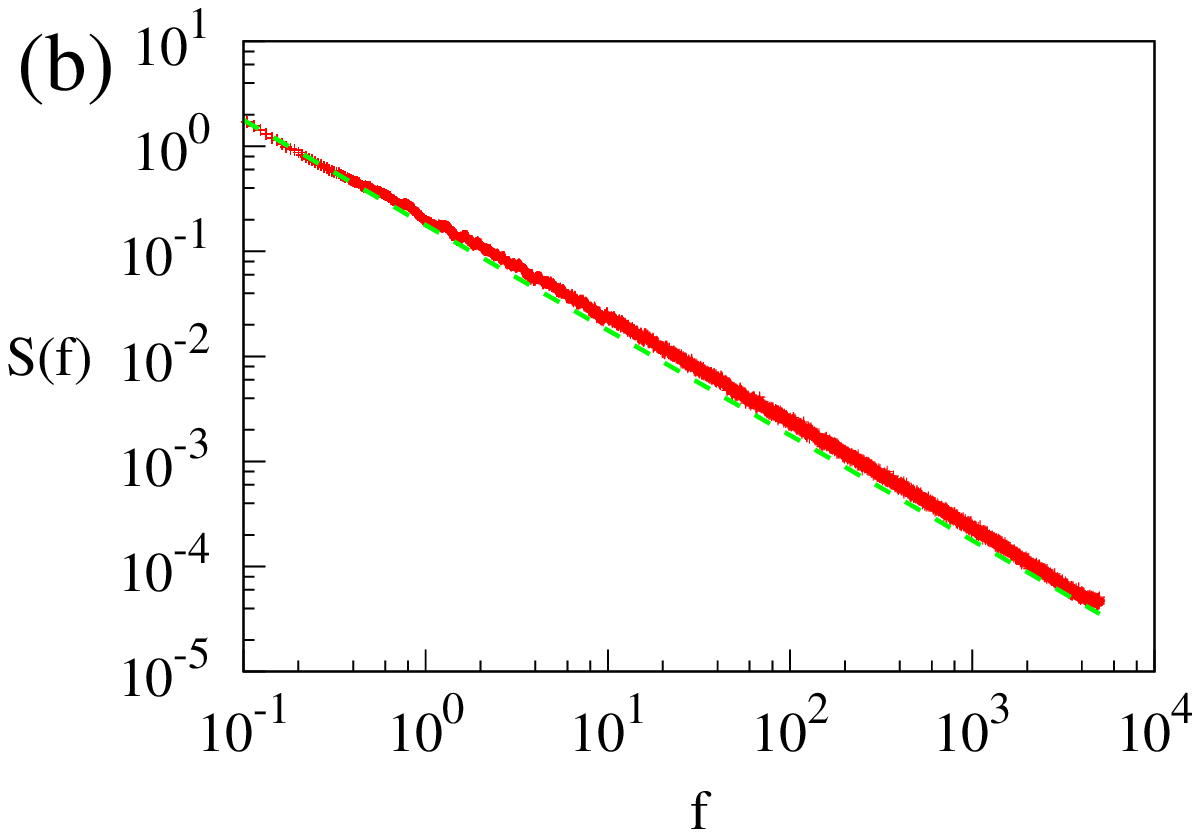}
\caption{(Color online) (a) Long-term PDF $P(x)$ of the signal generated by
Eqs.~(\ref{eq:exp}) and (\ref{eq:sde-mean-qexp}). The dashed (green) line is the
analytical expression (\ref{eq:pdf-q-exp-s}) for the long-term PDF. (b) Power
spectral density $S(f)$ of the same signal. The dashed (green) line shows the
slope $1/f$. The parameters used are $\lambda=3$, $\eta=5/2$, $x_0=1$,
$\sigma=1$, and $\bar{x}_{\mathrm{max}}=10^3$.}
\label{fig:super-q-exp}
\end{figure}

Comparison of numerically obtained long-term PDF and power spectral density
with analytical expressions is presented in Fig.~\ref{fig:super-q-exp}.
Numerical solution confirms the presence of the frequency region where the
power spectral density has $1/f^{\beta}$ dependence. In addition, the long-term
PDF of the signal deviates from the $q$-exponential function
(\ref{eq:pdf-q-exp-s}) only slightly.

\subsection{$q$-Gaussian distribution}

In order to obtain the $q$-Gaussian long-term PDF of the signal $x$ we will
consider the local stationary PDF conditioned to the value of the parameter
$\bar{x}$ in form of the Gaussian distribution, 
\begin{equation}
\varphi(x|\bar{x})=\frac{1}{\sqrt{\pi}\bar{x}}\exp(-x^2/\bar{x}^2)\,.
\label{eq:gauss}
\end{equation}
The standard deviation of $x$ in the distribution $\varphi(x|\bar{x})$ is
proportional to the parameter $\bar{x}$. The fluctuating parameter $\bar{x}$
obeys SDE with exponential restriction of diffusion (\ref{eq:sde-restricted})
with the parameter $m=2$, 
\begin{equation}
d\bar{x}=\sigma^2\left[\eta-\frac{\lambda}{2}+\frac{x_0^2}{\bar{x}^2}-\frac{
\bar{x}^2}{\bar{x}_{\mathrm{max}}^2}\right]\bar{x}^{2\eta-1}dt+\sigma\bar{x}^{
\eta}dW\,.
\label{eq:sde-mean-gauss}
\end{equation}
Here $x_0$ is the parameter describing exponential cut-off of the steady state
PDF of $\bar{x}$ at small values of $\bar{x}$, whereas the parameter
$\bar{x}_{\mathrm{max}}(\gg x_0)$ leads to the exponential cut-off at large
values of $\bar{x}$. When $\bar{x}\ll\bar{x}_{\mathrm{max}}$, the influence of
the exponential cut-off at large values of $\bar{x}$ is small. Neglecting
$\bar{x}_{\mathrm{max}}$ the steady state PDF from the Fokker-Planck equation
corresponding to Eq.~(\ref{eq:sde-mean-gauss}) is
\begin{equation}
p(\bar{x})=\frac{1}{x_0\Gamma\left(\frac{\lambda-1}{2}\right)}\left(\frac{x_0}{
\bar{x}}\right)^{\lambda}\exp\left(-\frac{x_0^2}{\bar{x}^2}\right)\,.
\label{eq:pdf-q-g}
\end{equation}
From Eqs.~(\ref{eq:super}), (\ref{eq:gauss}), and (\ref{eq:pdf-q-g}) we obtain
that for $x\ll\bar{x}_{\mathrm{max}}$ the long-term stationary PDF of the signal
$x$ is $q$-Gaussian, i.e., 
\begin{eqnarray}
P(x) & = &
\frac{\Gamma\left(\frac{\lambda}{2}\right)}{\sqrt{\pi}x_0\Gamma\left(\frac{
\lambda-1}{2}\right)}\left(\frac{x_0^2}{x_0^2+x^2}\right)^{\frac{\lambda}{
2}}=\frac{\Gamma\left(\frac{\lambda}{2}\right)}{\sqrt{\pi}x_0\Gamma\left(\frac{
\lambda-1}{2}\right)}\exp_q\left(-\lambda\frac{x^2}{2x_0^2}\right)\,,
\label{eq:pdf-q-gauss-s}
\\ q & = & 1+2/\lambda\,.\nonumber
\end{eqnarray}

\begin{figure}
\includegraphics[width=0.4\textwidth]{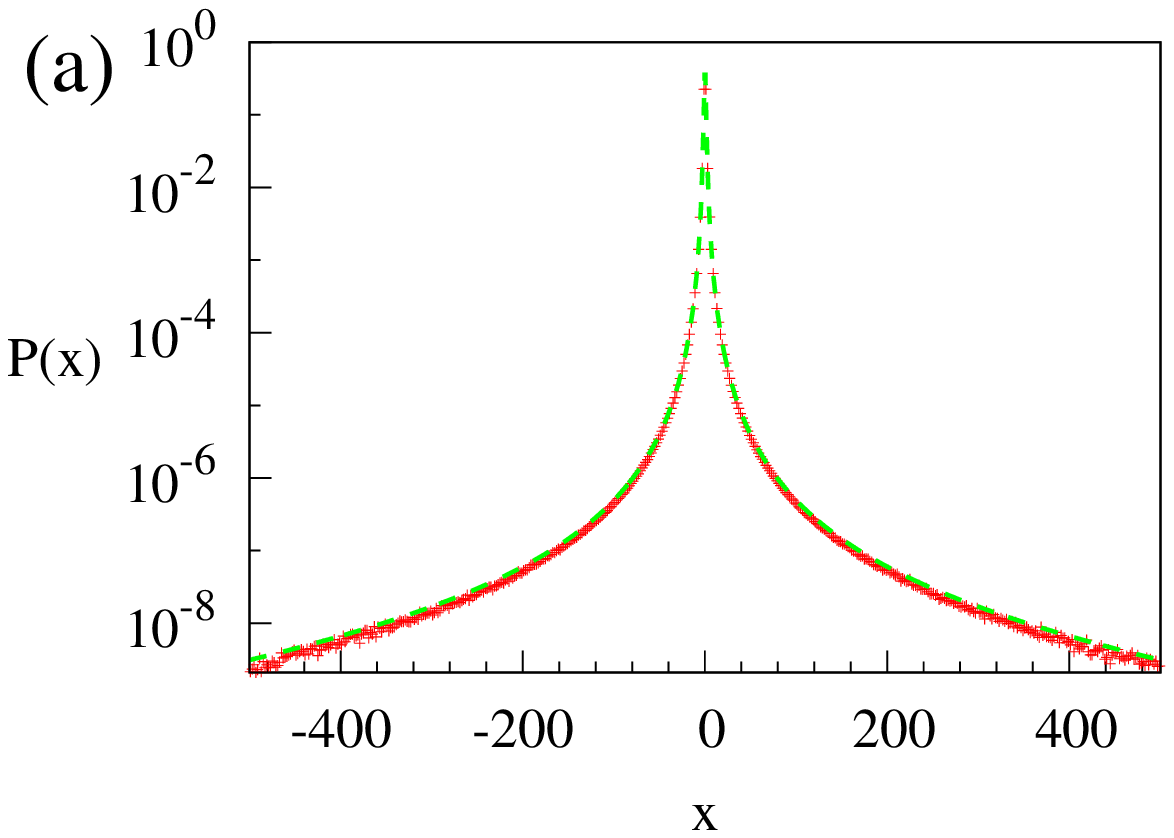}\includegraphics[width=0.4\textwidth]{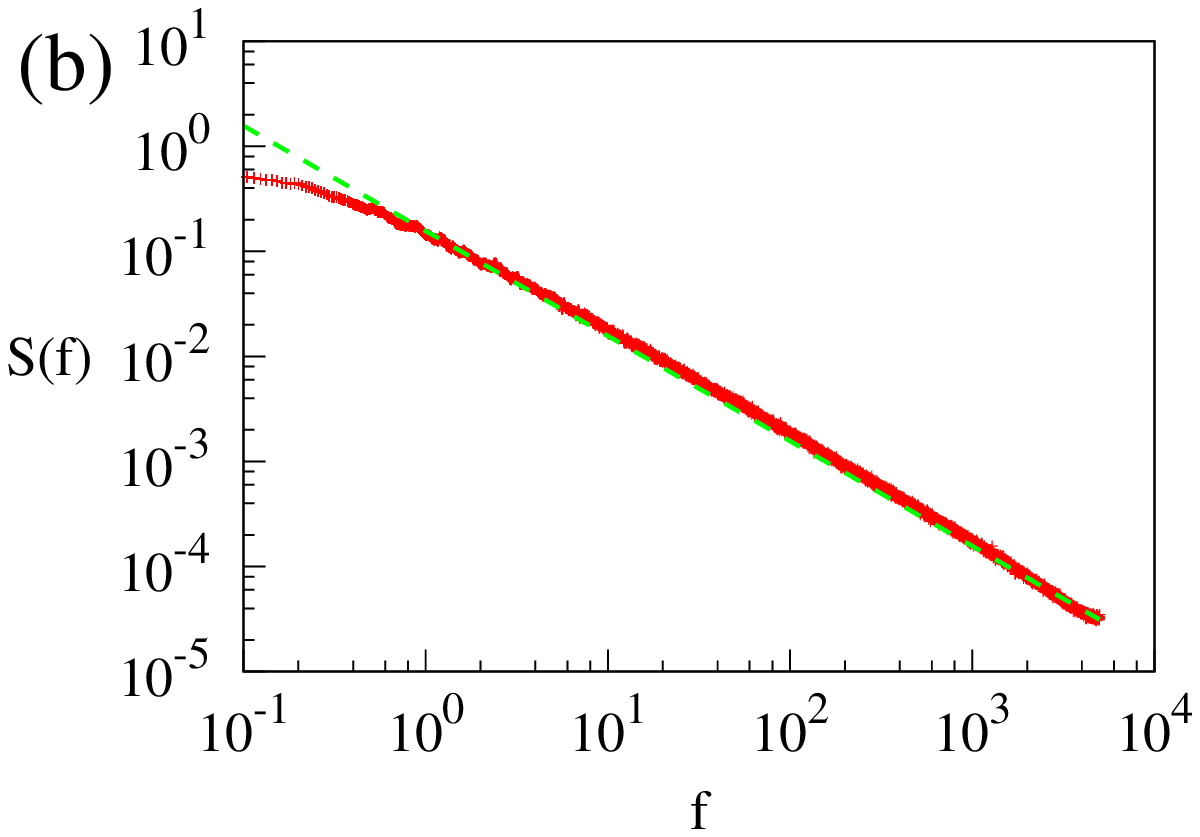}
\caption{(Color online) (a) Long-term PDF $P(x)$ of the signal generated by
Eqs.~(\ref{eq:gauss}) and (\ref{eq:sde-mean-gauss}). The dashed (green) line is
the analytical expression (\ref{eq:pdf-q-gauss-s}) for the long-term PDF. (b)
Power spectral density of the same signal. The dashed (green) line shows the
slope $1/f$. The parameters used are $\lambda=3$, $\eta=5/2$, $x_0=1$,
$\sigma=1$, and $\bar{x}_{\mathrm{max}}=10^3$.}
\label{fig:super-q-gauss}
\end{figure}

Comparison of numerically obtained long-term PDF and power spectral density with
analytical expressions is presented in Fig.~\ref{fig:super-q-gauss}. Numerical
solution confirms the presence of the frequency region where the power spectral
density has $1/f^{\beta}$ dependence. In addition, the long-term PDF of the
signal deviates only slightly from the $q$-Gaussian function
(\ref{eq:pdf-q-gauss-s}). In contrast to Sec.~\ref{sub:q-gauss}, the
superstatistical approach yields $1/f^{\beta}$ power spectral density only for
the absolute value $|x|$ of the signal. Since the signs of two consecutive
values of $x$ are uncorrelated, the spectrum of the signal $x$ itself in the
same frequency region is almost flat.

\section{Discussion}

Common characteristics of complex systems include long-range interactions,
long-range correlations, (multi)fractality, and non-Gaussian distributions with
asymptotic power-law behavior. The nonextensive statistical mechanics, which is
a generalization of the Boltzmann-Gibbs theory, is a possible theoretical
framework for describing these systems. However, long-range correlations and
$1/f^\beta$ noise has not been observed in previously used models giving
distributions of the nonextensive statistical mechanics. The joint reproduction of
the distributions of nonextensive statistical mechanics and $1/f$ noise,
presented in this paper, extends understanding of the complex systems. 

Modeling of the concrete systems by the nonlinear SDEs is not the goal of this
paper. However, relation of the models and obtained results with some features
of the financial systems can be pointed out. Equations with multiplicative noise
are already used for modeling of financial systems, e.g.\ the well-known $3/2$
model of stochastic volatility \cite{ahn-1999}. There is empirical evidence that
trading activity, trading volume, and volatility are stochastic variables with
the long-range correlation
\cite{Mantegna-2001,Lax06,Jeanblanc09,ahn-1999,engle-2001,plerou-2001,gabaix-2003}.
This key aspect, however, is not accounted for in the widely used models. On
the other hand, the empirical findings of PDF of the return and other
financial variables are successfully described within the nonextensive
statistical framework \cite{tsallis-2003a,drozdz-2010,gell-mann-2004}. The
return has a distribution that is very well fitted by $q$-Gaussians, only slowly
becoming Gaussian as the time scale approaches months, years and longer time
horizons. Another interesting statistic which can be modeled within the
nonextensive framework is the distribution of volumes, defined as the number of
shares traded. Modeling of some properties of the financial systems using the
point process models and SDEs have been undertaken in papers
\cite{gontis-2004,gontis-2009,gontis-2010,Kononovicius2011}. Equations presented
in this article incorporating the long-range correlations, $1/f^{\beta}$ noise,
and $q$-Gaussian distributions suggest deeper comprehension of these processes.


\begin{thebibliography}{10}%
\makeatletter
\providecommand \@ifxundefined [1]{%
 \ifx #1\undefined \expandafter \@firstoftwo
 \else \expandafter \@secondoftwo
\fi
}%
\providecommand \@ifnum [1]{%
 \ifnum #1\expandafter \@firstoftwo
 \else \expandafter \@secondoftwo
\fi
}%
\providecommand \enquote [1]{``#1''}%
\providecommand \bibnamefont  [1]{#1}%
\providecommand \bibfnamefont [1]{#1}%
\providecommand \citenamefont [1]{#1}%
\providecommand\href[0]{\@sanitize\@href}%
\providecommand\@href[1]{\endgroup\@@startlink{#1}\endgroup\@@href}%
\providecommand\@@href[1]{#1\@@endlink}%
\providecommand \@sanitize [0]{\begingroup\catcode`\&12\catcode`\#12\relax}%
\@ifxundefined \pdfoutput {\@firstoftwo}{%
 \@ifnum{\z@=\pdfoutput}{\@firstoftwo}{\@secondoftwo}%
}{%
 \providecommand\@@startlink[1]{\leavevmode}%
 \providecommand\@@endlink[0]{}%
}{%
 \providecommand\@@startlink[1]{%
  \leavevmode
  \pdfstartlink
   attr{/Border[0 0 1 ]/H/I/C[0 1 1]}%
   user{/Subtype/Link/A<</Type/Action/S/URI/URI(#1)>>}%
  \relax
 }%
 \providecommand\@@endlink[0]{\pdfendlink}%
}%
\providecommand \url  [0]{\begingroup\@sanitize \@url }%
\providecommand \@url [1]{\endgroup\@href {#1}{\urlprefix}}%
\providecommand \urlprefix [0]{URL }%
\providecommand \Eprint[0]{\href }%
\@ifxundefined \urlstyle {%
  \providecommand \doi [1]{doi:\discretionary{}{}{}#1}%
}{%
  \providecommand \doi [0]{doi:\discretionary{}{}{}\begingroup
  \urlstyle{rm}\Url }%
}%
\providecommand \doibase [0]{http://dx.doi.org/}%
\providecommand \Doi[1]{\href{\doibase#1}}%
\providecommand \bibAnnote [3]{%
  \BibitemShut{#1}%
  \begin{quotation}\noindent
    \textsc{Key:}\ #2\\\textsc{Annotation:}\ #3%
  \end{quotation}%
}%
\providecommand \bibAnnoteFile [2]{%
  \IfFileExists{#2}{\bibAnnote {#1} {#2} {\input{#2}}}{}%
}%
\providecommand \typeout [0]{\immediate \write \m@ne }%
\providecommand \selectlanguage [0]{\@gobble}%
\providecommand \bibinfo [0]{\@secondoftwo}%
\providecommand \bibfield [0]{\@secondoftwo}%
\providecommand \translation [1]{[#1]}%
\providecommand \BibitemOpen[0]{}%
\providecommand \bibitemStop [0]{}%
\providecommand \bibitemNoStop [0]{.\EOS\space}%
\providecommand \EOS [0]{\spacefactor3000\relax}%
\providecommand \BibitemShut [1]{\csname bibitem#1\endcsname}%
\bibitem{Mandelbrot-1999}%
  \BibitemOpen
  \bibfield{author}{%
  \bibinfo {author} {\bibfnamefont{B.~B.}\ \bibnamefont{Mandelbrot}},\ }%
  \emph{\bibinfo {title} {Multifractals and 1/f Noise: Wild Self-Affinity in
  Physics}}\ (\bibinfo {publisher} {Springer-Verlag},\ \bibinfo {address} {New
  York},\ \bibinfo {year} {1999})%
  \bibAnnoteFile{NoStop}{Mandelbrot-1999}%
\bibitem{Mantegna-2001}%
  \BibitemOpen
  \bibfield{author}{%
  \bibinfo {author} {\bibfnamefont{R.~N.}\ \bibnamefont{Mantegna}}\ and\
  \bibinfo {author} {\bibfnamefont{H.~E.}\ \bibnamefont{Stanley}},\ }%
  \emph{\bibinfo {title} {An Introduction to Econophysics: Correlations and
  Complexity}}\ (\bibinfo {publisher} {Cambridge University Press},\ \bibinfo
  {address} {Cambridge, UK},\ \bibinfo {year} {2001})%
  \bibAnnoteFile{NoStop}{Mantegna-2001}%
\bibitem{Lowen-2005}%
  \BibitemOpen
  \bibfield{author}{%
  \bibinfo {author} {\bibfnamefont{S.~B.}\ \bibnamefont{Lowen}}\ and\ \bibinfo
  {author} {\bibfnamefont{M.~C.}\ \bibnamefont{Teich}},\ }%
  \emph{\bibinfo {title} {Fractal-Based Point Processes}}\ (\bibinfo
  {publisher} {Wiley-Interscience},\ \bibinfo {address} {New Jersey},\ \bibinfo
  {year} {2005})%
  \bibAnnoteFile{NoStop}{Lowen-2005}%
\bibitem{tsallis-1988}%
  \BibitemOpen
  \bibfield{author}{%
  \bibinfo {author} {\bibfnamefont{C.}~\bibnamefont{Tsallis}},\ }%
  \bibfield{journal}{%
  \bibinfo {journal} {J. Stat. Phys.}\ }%
  \textbf{\bibinfo {volume} {52}},\ \bibinfo {pages} {479} (\bibinfo {year}
  {1988})%
  \bibAnnoteFile{NoStop}{tsallis-1988}%
\bibitem{queiros-2005-3}%
  \BibitemOpen
  \bibfield{author}{%
  \bibinfo {author} {\bibfnamefont{S.~M.~D.}\ \bibnamefont{Queiros}}, \bibinfo
  {author} {\bibfnamefont{C.}~\bibnamefont{Anteneodo}},\ and\ \bibinfo {author}
  {\bibfnamefont{C.}~\bibnamefont{Tsallis}},\ }%
  \bibfield{journal}{%
  \bibinfo {journal} {Proc. SPIE}\ }%
  \textbf{\bibinfo {volume} {5848}},\ \bibinfo {pages} {151} (\bibinfo {year}
  {2005})%
  \bibAnnoteFile{NoStop}{queiros-2005-3}%
\bibitem{Tsallis-2009}%
  \BibitemOpen
  \bibfield{author}{%
  \bibinfo {author} {\bibfnamefont{C.}~\bibnamefont{Tsallis}},\ }%
  \emph{\bibinfo {title} {Introduction to Nonextensive Statistical Mechanics --
  Approaching a Complex World}}\ (\bibinfo {publisher} {Springer},\ \bibinfo
  {address} {New York},\ \bibinfo {year} {2009})%
  \bibAnnoteFile{NoStop}{Tsallis-2009}%
\bibitem{tsallis-2009a}%
  \BibitemOpen
  \bibfield{author}{%
  \bibinfo {author} {\bibfnamefont{C.}~\bibnamefont{Tsallis}},\ }%
  \bibfield{journal}{%
  \bibinfo {journal} {Braz. J. Phys.}\ }%
  \textbf{\bibinfo {volume} {39}},\ \bibinfo {pages} {337} (\bibinfo {year}
  {2009})%
  \bibAnnoteFile{NoStop}{tsallis-2009a}%
\bibitem{telesca-2010}%
  \BibitemOpen
  \bibfield{author}{%
  \bibinfo {author} {\bibfnamefont{L.}~\bibnamefont{Telesca}},\ }%
  \bibfield{journal}{%
  \bibinfo {journal} {Tectonophysics}\ }%
  \textbf{\bibinfo {volume} {494}},\ \bibinfo {pages} {155} (\bibinfo {year}
  {2010})%
  \bibAnnoteFile{NoStop}{telesca-2010}%
\bibitem{kaulakys-2005}%
  \BibitemOpen
  \bibfield{author}{%
  \bibinfo {author} {\bibfnamefont{B.}~\bibnamefont{Kaulakys}}, \bibinfo
  {author} {\bibfnamefont{V.}~\bibnamefont{Gontis}},\ and\ \bibinfo {author}
  {\bibfnamefont{M.}~\bibnamefont{Alaburda}},\ }%
  \bibfield{journal}{%
  \bibinfo {journal} {Phys. Rev. E}\ }%
  \textbf{\bibinfo {volume} {71}},\ \bibinfo {pages} {051105} (\bibinfo {year}
  {2005})%
  \bibAnnoteFile{NoStop}{kaulakys-2005}%
\bibitem{kaulakys-2009}%
  \BibitemOpen
  \bibfield{author}{%
  \bibinfo {author} {\bibfnamefont{B.}~\bibnamefont{Kaulakys}}\ and\ \bibinfo
  {author} {\bibfnamefont{M.}~\bibnamefont{Alaburda}},\ }%
  \bibfield{journal}{%
  \bibinfo {journal} {J. Stat. Mech.}\ }%
  \textbf{\bibinfo {volume} {2009}},\ \bibinfo {pages} {P02051} (\bibinfo
  {year} {2009})%
  \bibAnnoteFile{NoStop}{kaulakys-2009}%
\bibitem{Fossion-2010}%
  \BibitemOpen
  \bibfield{author}{%
  \bibinfo {author} {\bibfnamefont{R.}~\bibnamefont{Fossion}}, \bibinfo
  {author} {\bibfnamefont{E.}~\bibnamefont{Landa}}, \bibinfo {author}
  {\bibfnamefont{P.}~\bibnamefont{Stransky}}, \bibinfo {author}
  {\bibfnamefont{V.}~\bibnamefont{Velazquez}}, \bibinfo {author}
  {\bibfnamefont{J.~C.~L.}\ \bibnamefont{Vieyra}}, \bibinfo {author}
  {\bibfnamefont{I.}~\bibnamefont{Garduno}}, \bibinfo {author}
  {\bibfnamefont{D.}~\bibnamefont{Garcia}},\ and\ \bibinfo {author}
  {\bibfnamefont{A.}~\bibnamefont{Frank}},\ }%
  \bibfield{journal}{%
  \bibinfo {journal} {AIP Conf. Proc.}\ }%
  \textbf{\bibinfo {volume} {1323}},\ \bibinfo {pages} {74} (\bibinfo {year}
  {2010})%
  \bibAnnoteFile{NoStop}{Fossion-2010}%
\bibitem{gell-mann-2004}%
  \BibitemOpen
  \bibfield{author}{%
  \bibinfo {author} {\bibfnamefont{C.~M.}\ \bibnamefont{Gell-Mann}}\ and\
  \bibinfo {author} {\bibfnamefont{C.}~\bibnamefont{Tsallis}},\ }%
  \emph{\bibinfo {title} {Nonextensive Entropy---Interdisciplinary
  Applications}}\ (\bibinfo {publisher} {Oxford University Press},\ \bibinfo
  {address} {NY},\ \bibinfo {year} {2004})%
  \bibAnnoteFile{NoStop}{gell-mann-2004}%
\bibitem{abe-2006}%
  \BibitemOpen
  \bibfield{author}{%
  \bibinfo {author} {\bibfnamefont{S.}~\bibnamefont{Abe}},\ }%
  \bibfield{journal}{%
  \bibinfo {journal} {Astrophys. Space Sci.}\ }%
  \textbf{\bibinfo {volume} {305}},\ \bibinfo {pages} {241} (\bibinfo {year}
  {2006})%
  \bibAnnoteFile{NoStop}{abe-2006}%
\bibitem{Picoli-2009}%
  \BibitemOpen
  \bibfield{author}{%
  \bibinfo {author} {\bibfnamefont{S.}~\bibnamefont{Picoli}}, \bibinfo {author}
  {\bibfnamefont{R.~S.}\ \bibnamefont{Mendes}}, \bibinfo {author}
  {\bibfnamefont{L.~C.}\ \bibnamefont{Malacarne}},\ and\ \bibinfo {author}
  {\bibfnamefont{R.~P.~B.}\ \bibnamefont{Santos}},\ }%
  \bibfield{journal}{%
  \bibinfo {journal} {Braz. J. Phys.}\ }%
  \textbf{\bibinfo {volume} {39}},\ \bibinfo {pages} {468} (\bibinfo {year}
  {2009})%
  \bibAnnoteFile{NoStop}{Picoli-2009}%
\bibitem{tsallis-1998}%
  \BibitemOpen
  \bibfield{author}{%
  \bibinfo {author} {\bibfnamefont{C.}~\bibnamefont{Tsallis}}, \bibinfo
  {author} {\bibfnamefont{A.~R.}\ \bibnamefont{Plastino}},\ and\ \bibinfo
  {author} {\bibfnamefont{R.~S.}\ \bibnamefont{Mendes}},\ }%
  \bibfield{journal}{%
  \bibinfo {journal} {Physica A}\ }%
  \textbf{\bibinfo {volume} {261}},\ \bibinfo {pages} {534} (\bibinfo {year}
  {1998})%
  \bibAnnoteFile{NoStop}{tsallis-1998}%
\bibitem{prato-1999}%
  \BibitemOpen
  \bibfield{author}{%
  \bibinfo {author} {\bibfnamefont{D.}~\bibnamefont{Prato}}\ and\ \bibinfo
  {author} {\bibfnamefont{C.}~\bibnamefont{Tsallis}},\ }%
  \bibfield{journal}{%
  \bibinfo {journal} {Phys. Rev. E}\ }%
  \textbf{\bibinfo {volume} {60}},\ \bibinfo {pages} {2398} (\bibinfo {year}
  {1999})%
  \bibAnnoteFile{NoStop}{prato-1999}%
\bibitem{tsallis-1999}%
  \BibitemOpen
  \bibfield{author}{%
  \bibinfo {author} {\bibfnamefont{C.}~\bibnamefont{Tsallis}},\ }%
  \bibfield{journal}{%
  \bibinfo {journal} {Braz. J. Phys.}\ }%
  \textbf{\bibinfo {volume} {29}},\ \bibinfo {pages} {1} (\bibinfo {year}
  {1999})%
  \bibAnnoteFile{NoStop}{tsallis-1999}%
\bibitem{hanel-2011a}%
  \BibitemOpen
  \bibfield{author}{%
  \bibinfo {author} {\bibfnamefont{R.}~\bibnamefont{Hanel}}\ and\ \bibinfo
  {author} {\bibfnamefont{S.}~\bibnamefont{Thurner}},\ }%
  \bibfield{journal}{%
  \bibinfo {journal} {EPL}\ }%
  \textbf{\bibinfo {volume} {93}},\ \bibinfo {pages} {20006} (\bibinfo {year}
  {2011})%
  \bibAnnoteFile{NoStop}{hanel-2011a}%
\bibitem{hanel-2011b}%
  \BibitemOpen
  \bibfield{author}{%
  \bibinfo {author} {\bibfnamefont{R.}~\bibnamefont{Hanel}}, \bibinfo {author}
  {\bibfnamefont{S.}~\bibnamefont{Thurner}},\ and\ \bibinfo {author}
  {\bibfnamefont{M.}~\bibnamefont{Gell-Mann}},\ }%
  \bibfield{journal}{%
  \bibinfo {journal} {PNAS}\ }%
  \textbf{\bibinfo {volume} {108}},\ \bibinfo {pages} {6390} (\bibinfo {year}
  {2011})%
  \bibAnnoteFile{NoStop}{hanel-2011b}%
\bibitem{lyra-1998}%
  \BibitemOpen
  \bibfield{author}{%
  \bibinfo {author} {\bibfnamefont{M.~L.}\ \bibnamefont{Lyra}}\ and\ \bibinfo
  {author} {\bibfnamefont{C.}~\bibnamefont{Tsallis}},\ }%
  \bibfield{journal}{%
  \bibinfo {journal} {Phys. Rev. Lett.}\ }%
  \textbf{\bibinfo {volume} {80}},\ \bibinfo {pages} {53} (\bibinfo {year}
  {1998})%
  \bibAnnoteFile{NoStop}{lyra-1998}%
\bibitem{baldovin-2000}%
  \BibitemOpen
  \bibfield{author}{%
  \bibinfo {author} {\bibfnamefont{F.}~\bibnamefont{Baldovin}}\ and\ \bibinfo
  {author} {\bibfnamefont{A.}~\bibnamefont{Robledo}},\ }%
  \bibfield{journal}{%
  \bibinfo {journal} {Europhys. Lett.}\ }%
  \textbf{\bibinfo {volume} {60}},\ \bibinfo {pages} {518} (\bibinfo {year}
  {2000})%
  \bibAnnoteFile{NoStop}{baldovin-2000}%
\bibitem{borland-1998}%
  \BibitemOpen
  \bibfield{author}{%
  \bibinfo {author} {\bibfnamefont{L.}~\bibnamefont{Borland}},\ }%
  \bibfield{journal}{%
  \bibinfo {journal} {Phys. Rev. E}\ }%
  \textbf{\bibinfo {volume} {57}},\ \bibinfo {pages} {6634} (\bibinfo {year}
  {1998})%
  \bibAnnoteFile{NoStop}{borland-1998}%
\bibitem{beck-2001}%
  \BibitemOpen
  \bibfield{author}{%
  \bibinfo {author} {\bibfnamefont{C.}~\bibnamefont{Beck}}, \bibinfo {author}
  {\bibfnamefont{G.~S.}\ \bibnamefont{Lewis}},\ and\ \bibinfo {author}
  {\bibfnamefont{H.~L.}\ \bibnamefont{Swinney}},\ }%
  \bibfield{journal}{%
  \bibinfo {journal} {Phys. Rev. E}\ }%
  \textbf{\bibinfo {volume} {63}},\ \bibinfo {pages} {035303(R)} (\bibinfo
  {year} {2001})%
  \bibAnnoteFile{NoStop}{beck-2001}%
\bibitem{beck-2003}%
  \BibitemOpen
  \bibfield{author}{%
  \bibinfo {author} {\bibfnamefont{C.}~\bibnamefont{Beck}}\ and\ \bibinfo
  {author} {\bibfnamefont{E.~G.}\ \bibnamefont{Cohen}},\ }%
  \bibfield{journal}{%
  \bibinfo {journal} {Physica A}\ }%
  \textbf{\bibinfo {volume} {322}},\ \bibinfo {pages} {267} (\bibinfo {year}
  {2003})%
  \bibAnnoteFile{NoStop}{beck-2003}%
\bibitem{wilk-2000}%
  \BibitemOpen
  \bibfield{author}{%
  \bibinfo {author} {\bibfnamefont{G.}~\bibnamefont{Wilk}}\ and\ \bibinfo
  {author} {\bibfnamefont{Z.}~\bibnamefont{W\l{}odarczyk}},\ }%
  \bibfield{journal}{%
  \bibinfo {journal} {Phys. Rev. Lett.}\ }%
  \textbf{\bibinfo {volume} {84}},\ \bibinfo {pages} {2770} (\bibinfo {year}
  {2000})%
  \bibAnnoteFile{NoStop}{wilk-2000}%
\bibitem{beck-2001-1}%
  \BibitemOpen
  \bibfield{author}{%
  \bibinfo {author} {\bibfnamefont{C.}~\bibnamefont{Beck}},\ }%
  \bibfield{journal}{%
  \bibinfo {journal} {Phys. Rev. Lett.}\ }%
  \textbf{\bibinfo {volume} {87}},\ \bibinfo {pages} {180601} (\bibinfo {year}
  {2001})%
  \bibAnnoteFile{NoStop}{beck-2001-1}%
\bibitem{latora-2001}%
  \BibitemOpen
  \bibfield{author}{%
  \bibinfo {author} {\bibfnamefont{V.}~\bibnamefont{Latora}}, \bibinfo {author}
  {\bibfnamefont{A.}~\bibnamefont{Rapisarda}},\ and\ \bibinfo {author}
  {\bibfnamefont{C.}~\bibnamefont{Tsallis}},\ }%
  \bibfield{journal}{%
  \bibinfo {journal} {Phys. Rev. E}\ }%
  \textbf{\bibinfo {volume} {64}},\ \bibinfo {pages} {056134} (\bibinfo {year}
  {2001})%
  \bibAnnoteFile{NoStop}{latora-2001}%
\bibitem{tsallis-2003a}%
  \BibitemOpen
  \bibfield{author}{%
  \bibinfo {author} {\bibfnamefont{C.}~\bibnamefont{Tsallis}}, \bibinfo
  {author} {\bibfnamefont{C.}~\bibnamefont{Anteneodo}}, \bibinfo {author}
  {\bibfnamefont{L.}~\bibnamefont{Borland}},\ and\ \bibinfo {author}
  {\bibfnamefont{R.}~\bibnamefont{Osorio}},\ }%
  \bibfield{journal}{%
  \bibinfo {journal} {Physica A}\ }%
  \textbf{\bibinfo {volume} {324}},\ \bibinfo {pages} {89} (\bibinfo {year}
  {2003})%
  \bibAnnoteFile{NoStop}{tsallis-2003a}%
\bibitem{drozdz-2010}%
  \BibitemOpen
  \bibfield{author}{%
  \bibinfo {author} {\bibfnamefont{S.}~\bibnamefont{Drozdz}}, \bibinfo {author}
  {\bibfnamefont{J.}~\bibnamefont{Kwapien}}, \bibinfo {author}
  {\bibfnamefont{P.}~\bibnamefont{Oswiecimka}},\ and\ \bibinfo {author}
  {\bibfnamefont{R.}~\bibnamefont{Rak}},\ }%
  \bibfield{journal}{%
  \bibinfo {journal} {New J. Phys.}\ }%
  \textbf{\bibinfo {volume} {12}},\ \bibinfo {pages} {105003} (\bibinfo {year}
  {2010})%
  \bibAnnoteFile{NoStop}{drozdz-2010}%
\bibitem{borland-2002}%
  \BibitemOpen
  \bibfield{author}{%
  \bibinfo {author} {\bibfnamefont{L.}~\bibnamefont{Borland}},\ }%
  \bibfield{journal}{%
  \bibinfo {journal} {Phys. Rev. Lett.}\ }%
  \textbf{\bibinfo {volume} {89}},\ \bibinfo {pages} {098701} (\bibinfo {year}
  {2002})%
  \bibAnnoteFile{NoStop}{borland-2002}%
\bibitem{anteneodo-2003}%
  \BibitemOpen
  \bibfield{author}{%
  \bibinfo {author} {\bibfnamefont{C.}~\bibnamefont{Anteneodo}}\ and\ \bibinfo
  {author} {\bibfnamefont{C.}~\bibnamefont{Tsallis}},\ }%
  \bibfield{journal}{%
  \bibinfo {journal} {J. Math. Phys.}\ }%
  \textbf{\bibinfo {volume} {72}},\ \bibinfo {pages} {5194} (\bibinfo {year}
  {2003})%
  \bibAnnoteFile{NoStop}{anteneodo-2003}%
\bibitem{santos-2010}%
  \BibitemOpen
  \bibfield{author}{%
  \bibinfo {author} {\bibfnamefont{B.}~\bibnamefont{Coutinho~dos Santos}}\ and\
  \bibinfo {author} {\bibfnamefont{C.}~\bibnamefont{Tsallis}},\ }%
  \bibfield{journal}{%
  \bibinfo {journal} {Phys. Rev. E}\ }%
  \textbf{\bibinfo {volume} {82}},\ \bibinfo {pages} {061119} (\bibinfo {year}
  {2010})%
  \bibAnnoteFile{NoStop}{santos-2010}%
\bibitem{queiros-2007}%
  \BibitemOpen
  \bibfield{author}{%
  \bibinfo {author} {\bibfnamefont{S.~M.~D.}\ \bibnamefont{Queiros}}, \bibinfo
  {author} {\bibfnamefont{L.~G.}\ \bibnamefont{Moyano}}, \bibinfo {author}
  {\bibfnamefont{J.}~\bibnamefont{de~Souza}},\ and\ \bibinfo {author}
  {\bibfnamefont{C.}~\bibnamefont{Tsallis}},\ }%
  \bibfield{journal}{%
  \bibinfo {journal} {Eur. Phys. J. B}\ }%
  \textbf{\bibinfo {volume} {55}},\ \bibinfo {pages} {161} (\bibinfo {year}
  {2007})%
  \bibAnnoteFile{NoStop}{queiros-2007}%
\bibitem{scholarpedia-2009}%
  \BibitemOpen
  \bibfield{author}{%
  \bibinfo {author} {\bibfnamefont{L.~M.}\ \bibnamefont{Ward}}\ and\ \bibinfo
  {author} {\bibfnamefont{P.~E.}\ \bibnamefont{Greenwood}},\ }%
  \bibfield{journal}{%
  \bibinfo {journal} {Scholarpedia}\ }%
  \textbf{\bibinfo {volume} {2}},\ \bibinfo {pages} {1537} (\bibinfo {year}
  {2007})%
  \bibAnnoteFile{NoStop}{scholarpedia-2009}%
\bibitem{weissman-1988}%
  \BibitemOpen
  \bibfield{author}{%
  \bibinfo {author} {\bibfnamefont{M.~B.}\ \bibnamefont{Weissman}},\ }%
  \bibfield{journal}{%
  \bibinfo {journal} {Rev. Mod. Phys.}\ }%
  \textbf{\bibinfo {volume} {60}},\ \bibinfo {pages} {537} (\bibinfo {year}
  {1988})%
  \bibAnnoteFile{NoStop}{weissman-1988}%
\bibitem{Gilden-1995}%
  \BibitemOpen
  \bibfield{author}{%
  \bibinfo {author} {\bibfnamefont{D.~L.}\ \bibnamefont{Gilden}}, \bibinfo
  {author} {\bibfnamefont{T.}~\bibnamefont{Thornton}},\ and\ \bibinfo {author}
  {\bibfnamefont{M.~W.}\ \bibnamefont{Mallon}},\ }%
  \bibfield{journal}{%
  \bibinfo {journal} {Science}\ }%
  \textbf{\bibinfo {volume} {267}},\ \bibinfo {pages} {1837} (\bibinfo {year}
  {1995})%
  \bibAnnoteFile{NoStop}{Gilden-1995}%
\bibitem{milotti-2002}%
  \BibitemOpen
  \bibfield{author}{%
  \bibinfo {author} {\bibfnamefont{E.}~\bibnamefont{Milotti}}}%
   (\bibinfo {year} {2002}),\
  \Eprint{http://arxiv.org/abs/arXiv:physics/0204033v1
  [physics.class-ph]}{arXiv:physics/0204033v1 [physics.class-ph]}%
  \bibAnnoteFile{NoStop}{milotti-2002}%
\bibitem{wong-2003}%
  \BibitemOpen
  \bibfield{author}{%
  \bibinfo {author} {\bibfnamefont{H.}~\bibnamefont{Wong}},\ }%
  \bibfield{journal}{%
  \bibinfo {journal} {Microelectron. Reliab.}\ }%
  \textbf{\bibinfo {volume} {43}},\ \bibinfo {pages} {585} (\bibinfo {year}
  {2003})%
  \bibAnnoteFile{NoStop}{wong-2003}%
\bibitem{Gardiner04}%
  \BibitemOpen
  \bibfield{author}{%
  \bibinfo {author} {\bibfnamefont{C.~W.}\ \bibnamefont{Gardiner}},\ }%
  \emph{\bibinfo {title} {Handbook of Stochastic Methods for Physics, Chemistry
  and the Natural Sciences}}\ (\bibinfo {publisher} {Springer-Verlag},\
  \bibinfo {address} {Berlin},\ \bibinfo {year} {2004})%
  \bibAnnoteFile{NoStop}{Gardiner04}%
\bibitem{Risken89}%
  \BibitemOpen
  \bibfield{author}{%
  \bibinfo {author} {\bibfnamefont{H.}~\bibnamefont{Risken}},\ }%
  \emph{\bibinfo {title} {The Fokker-Planck Equation: Methods of Solution and
  Applications}}\ (\bibinfo {publisher} {Springer-Verlag},\ \bibinfo {address}
  {Berlin},\ \bibinfo {year} {1989})%
  \bibAnnoteFile{NoStop}{Risken89}%
\bibitem{Farias09}%
  \BibitemOpen
  \bibfield{author}{%
  \bibinfo {author} {\bibfnamefont{R.~L.~S.}\ \bibnamefont{Farias}}, \bibinfo
  {author} {\bibfnamefont{R.~O.}\ \bibnamefont{Ramos}},\ and\ \bibinfo {author}
  {\bibfnamefont{L.~A.}\ \bibnamefont{da~Silva}},\ }%
  \bibfield{journal}{%
  \bibinfo {journal} {Phys. Rev. E}\ }%
  \textbf{\bibinfo {volume} {80}},\ \bibinfo {pages} {031143} (\bibinfo {year}
  {2009})%
  \bibAnnoteFile{NoStop}{Farias09}%
\bibitem{Lax06}%
  \BibitemOpen
  \bibfield{author}{%
  \bibinfo {author} {\bibfnamefont{M.}~\bibnamefont{Lax}}, \bibinfo {author}
  {\bibfnamefont{W.}~\bibnamefont{Cai}},\ and\ \bibinfo {author}
  {\bibfnamefont{M.}~\bibnamefont{Xu}},\ }%
  \emph{\bibinfo {title} {Random Processes in Physics and Finance}}\ (\bibinfo
  {publisher} {Oxford University Press},\ \bibinfo {address} {New York},\
  \bibinfo {year} {2006})%
  \bibAnnoteFile{NoStop}{Lax06}%
\bibitem{Jeanblanc09}%
  \BibitemOpen
  \bibfield{author}{%
  \bibinfo {author} {\bibfnamefont{M.}~\bibnamefont{Jeanblanc}}, \bibinfo
  {author} {\bibfnamefont{M.}~\bibnamefont{Yor}},\ and\ \bibinfo {author}
  {\bibfnamefont{M.}~\bibnamefont{Chesney}},\ }%
  \emph{\bibinfo {title} {Mathematical Methods for Financial Markets}}\
  (\bibinfo {publisher} {Springer},\ \bibinfo {address} {London},\ \bibinfo
  {year} {2009})%
  \bibAnnoteFile{NoStop}{Jeanblanc09}%
\bibitem{kaulakys-1998}%
  \BibitemOpen
  \bibfield{author}{%
  \bibinfo {author} {\bibfnamefont{B.}~\bibnamefont{Kaulakys}}\ and\ \bibinfo
  {author} {\bibfnamefont{T.}~\bibnamefont{Me\v{s}kauskas}},\ }%
  \bibfield{journal}{%
  \bibinfo {journal} {Phys. Rev. E}\ }%
  \textbf{\bibinfo {volume} {58}},\ \bibinfo {pages} {7013} (\bibinfo {year}
  {1998})%
  \bibAnnoteFile{NoStop}{kaulakys-1998}%
\bibitem{kaulakys-1999}%
  \BibitemOpen
  \bibfield{author}{%
  \bibinfo {author} {\bibfnamefont{B.}~\bibnamefont{Kaulakys}},\ }%
  \bibfield{journal}{%
  \bibinfo {journal} {Phys. Lett. A}\ }%
  \textbf{\bibinfo {volume} {257}},\ \bibinfo {pages} {37} (\bibinfo {year}
  {1999})%
  \bibAnnoteFile{NoStop}{kaulakys-1999}%
\bibitem{kaulakys-2001}%
  \BibitemOpen
  \bibfield{author}{%
  \bibinfo {author} {\bibfnamefont{B.}~\bibnamefont{Kaulakys}}\ and\ \bibinfo
  {author} {\bibfnamefont{T.}~\bibnamefont{Me\v{s}kauskas}},\ }%
  \bibfield{journal}{%
  \bibinfo {journal} {Microel. Reliab.}\ }%
  \textbf{\bibinfo {volume} {40}},\ \bibinfo {pages} {1781} (\bibinfo {year}
  {2000})%
  \bibAnnoteFile{NoStop}{kaulakys-2001}%
\bibitem{kaulakys-2002}%
  \BibitemOpen
  \bibfield{author}{%
  \bibinfo {author} {\bibfnamefont{B.}~\bibnamefont{Kaulakys}},\ }%
  \bibfield{journal}{%
  \bibinfo {journal} {Microel. Reliab.}\ }%
  \textbf{\bibinfo {volume} {40}},\ \bibinfo {pages} {1787} (\bibinfo {year}
  {2000})%
  \bibAnnoteFile{NoStop}{kaulakys-2002}%
\bibitem{kaulakys-2003}%
  \BibitemOpen
  \bibfield{author}{%
  \bibinfo {author} {\bibfnamefont{B.}~\bibnamefont{Kaulakys}},\ }%
  \bibfield{journal}{%
  \bibinfo {journal} {Lithuanian J. Phys.}\ }%
  \textbf{\bibinfo {volume} {40}},\ \bibinfo {pages} {281} (\bibinfo {year}
  {2000})%
  \bibAnnoteFile{NoStop}{kaulakys-2003}%
\bibitem{kaulakys-2004}%
  \BibitemOpen
  \bibfield{author}{%
  \bibinfo {author} {\bibfnamefont{B.}~\bibnamefont{Kaulakys}}\ and\ \bibinfo
  {author} {\bibfnamefont{J.}~\bibnamefont{Ruseckas}},\ }%
  \bibfield{journal}{%
  \bibinfo {journal} {Phys. Rev. E}\ }%
  \textbf{\bibinfo {volume} {70}},\ \bibinfo {pages} {020101(R)} (\bibinfo
  {year} {2004})%
  \bibAnnoteFile{NoStop}{kaulakys-2004}%
\bibitem{kaulakys-2006}%
  \BibitemOpen
  \bibfield{author}{%
  \bibinfo {author} {\bibfnamefont{B.}~\bibnamefont{Kaulakys}}, \bibinfo
  {author} {\bibfnamefont{J.}~\bibnamefont{Ruseckas}}, \bibinfo {author}
  {\bibfnamefont{V.}~\bibnamefont{Gontis}},\ and\ \bibinfo {author}
  {\bibfnamefont{M.}~\bibnamefont{Alaburda}},\ }%
  \bibfield{journal}{%
  \bibinfo {journal} {Physica A}\ }%
  \textbf{\bibinfo {volume} {365}},\ \bibinfo {pages} {217} (\bibinfo {year}
  {2006})%
  \bibAnnoteFile{NoStop}{kaulakys-2006}%
\bibitem{arnold-2000}%
  \BibitemOpen
  \bibfield{author}{%
  \bibinfo {author} {\bibfnamefont{P.}~\bibnamefont{Arnold}},\ }%
  \bibfield{journal}{%
  \bibinfo {journal} {Phys. Rev. E}\ }%
  \textbf{\bibinfo {volume} {61}},\ \bibinfo {pages} {6091} (\bibinfo {year}
  {2000})%
  \bibAnnoteFile{NoStop}{arnold-2000}%
\bibitem{gontis-2004}%
  \BibitemOpen
  \bibfield{author}{%
  \bibinfo {author} {\bibfnamefont{V.}~\bibnamefont{Gontis}}\ and\ \bibinfo
  {author} {\bibfnamefont{B.}~\bibnamefont{Kaulakys}},\ }%
  \bibfield{journal}{%
  \bibinfo {journal} {Physica A}\ }%
  \textbf{\bibinfo {volume} {343}},\ \bibinfo {pages} {505} (\bibinfo {year}
  {2004})%
  \bibAnnoteFile{NoStop}{gontis-2004}%
\bibitem{kaulakys-2009-3}%
  \BibitemOpen
  \bibfield{author}{%
  \bibinfo {author} {\bibfnamefont{B.}~\bibnamefont{Kaulakys}}, \bibinfo
  {author} {\bibfnamefont{M.}~\bibnamefont{Alaburda}}, \bibinfo {author}
  {\bibfnamefont{V.}~\bibnamefont{Gontis}},\ and\ \bibinfo {author}
  {\bibfnamefont{J.}~\bibnamefont{Ruseckas}},\ }%
  \bibfield{journal}{%
  \bibinfo {journal} {Braz. J. Phys.}\ }%
  \textbf{\bibinfo {volume} {39}},\ \bibinfo {pages} {453} (\bibinfo {year}
  {2009})%
  \bibAnnoteFile{NoStop}{kaulakys-2009-3}%
\bibitem{mamontov-1997}%
  \BibitemOpen
  \bibfield{author}{%
  \bibinfo {author} {\bibfnamefont{Y.~V.}\ \bibnamefont{Mamontov}}\ and\
  \bibinfo {author} {\bibfnamefont{M.}~\bibnamefont{Willander}},\ }%
  \bibfield{journal}{%
  \bibinfo {journal} {Nonlinear Dyn.}\ }%
  \textbf{\bibinfo {volume} {12}},\ \bibinfo {pages} {399} (\bibinfo {year}
  {1997})%
  \bibAnnoteFile{NoStop}{mamontov-1997}%
\bibitem{ruseckas-2010}%
  \BibitemOpen
  \bibfield{author}{%
  \bibinfo {author} {\bibfnamefont{J.}~\bibnamefont{Ruseckas}}\ and\ \bibinfo
  {author} {\bibfnamefont{B.}~\bibnamefont{Kaulakys}},\ }%
  \bibfield{journal}{%
  \bibinfo {journal} {Phys. Rev. E}\ }%
  \textbf{\bibinfo {volume} {81}},\ \bibinfo {pages} {031105} (\bibinfo {year}
  {2010})%
  \bibAnnoteFile{NoStop}{ruseckas-2010}%
\bibitem{Erland2011}%
  \BibitemOpen
  \bibfield{author}{%
  \bibinfo {author} {\bibfnamefont{S.}~\bibnamefont{Erland}}, \bibinfo {author}
  {\bibfnamefont{P.~E.}\ \bibnamefont{Greenwood}},\ and\ \bibinfo {author}
  {\bibfnamefont{L.~M.}\ \bibnamefont{Ward}},\ }%
  \bibfield{journal}{%
  \bibinfo {journal} {EPL (Europhysics Letters)}\ }%
  \textbf{\bibinfo {volume} {95}},\ \bibinfo {pages} {60006} (\bibinfo {year}
  {2011})%
  \bibAnnoteFile{NoStop}{Erland2011}%
\bibitem{kaulakys-2009-2}%
  \BibitemOpen
  \bibfield{author}{%
  \bibinfo {author} {\bibfnamefont{B.}~\bibnamefont{Kaulakys}}, \bibinfo
  {author} {\bibfnamefont{M.}~\bibnamefont{Alaburda}},\ and\ \bibinfo {author}
  {\bibfnamefont{V.}~\bibnamefont{Gontis}},\ }%
  \bibfield{journal}{%
  \bibinfo {journal} {AIP Conf. Proc.}\ }%
  \textbf{\bibinfo {volume} {1129}},\ \bibinfo {pages} {13} (\bibinfo {year}
  {2009})%
  \bibAnnoteFile{NoStop}{kaulakys-2009-2}%
\bibitem{gontis-2009}%
  \BibitemOpen
  \bibfield{author}{%
  \bibinfo {author} {\bibfnamefont{V.}~\bibnamefont{Gontis}}, \bibinfo {author}
  {\bibfnamefont{B.}~\bibnamefont{Kaulakys}},\ and\ \bibinfo {author}
  {\bibfnamefont{J.}~\bibnamefont{Ruseckas}},\ }%
  \bibfield{journal}{%
  \bibinfo {journal} {AIP Conf. Proc.}\ }%
  \textbf{\bibinfo {volume} {1129}},\ \bibinfo {pages} {563} (\bibinfo {year}
  {2009})%
  \bibAnnoteFile{NoStop}{gontis-2009}%
\bibitem{gontis-2010}%
  \BibitemOpen
  \bibfield{author}{%
  \bibinfo {author} {\bibfnamefont{V.}~\bibnamefont{Gontis}}, \bibinfo {author}
  {\bibfnamefont{J.}~\bibnamefont{Ruseckas}},\ and\ \bibinfo {author}
  {\bibfnamefont{A.}~\bibnamefont{Kononovi\v{c}ius}},\ }%
  \bibfield{journal}{%
  \bibinfo {journal} {Physica A}\ }%
  \textbf{\bibinfo {volume} {389}},\ \bibinfo {pages} {100} (\bibinfo {year}
  {2010})%
  \bibAnnoteFile{NoStop}{gontis-2010}%
\bibitem{tsallis-2003}%
  \BibitemOpen
  \bibfield{author}{%
  \bibinfo {author} {\bibfnamefont{C.}~\bibnamefont{Tsallis}}\ and\ \bibinfo
  {author} {\bibfnamefont{A.~M.~C.}\ \bibnamefont{Souza}},\ }%
  \bibfield{journal}{%
  \bibinfo {journal} {Phys. Rev. E}\ }%
  \textbf{\bibinfo {volume} {67}},\ \bibinfo {pages} {026106} (\bibinfo {year}
  {2003})%
  \bibAnnoteFile{NoStop}{tsallis-2003}%
\bibitem{abe-2007}%
  \BibitemOpen
  \bibfield{author}{%
  \bibinfo {author} {\bibfnamefont{S.}~\bibnamefont{Abe}}, \bibinfo {author}
  {\bibfnamefont{C.}~\bibnamefont{Beck}},\ and\ \bibinfo {author}
  {\bibfnamefont{E.~G.~D.}\ \bibnamefont{Cohen}},\ }%
  \bibfield{journal}{%
  \bibinfo {journal} {Phys. Rev. E}\ }%
  \textbf{\bibinfo {volume} {76}},\ \bibinfo {pages} {031102} (\bibinfo {year}
  {2007})%
  \bibAnnoteFile{NoStop}{abe-2007}%
\bibitem{hahn-2010}%
  \BibitemOpen
  \bibfield{author}{%
  \bibinfo {author} {\bibfnamefont{M.~G.}\ \bibnamefont{Hahn}}, \bibinfo
  {author} {\bibfnamefont{X.}~\bibnamefont{Jiang}},\ and\ \bibinfo {author}
  {\bibfnamefont{S.}~\bibnamefont{Umarov}},\ }%
  \bibfield{journal}{%
  \bibinfo {journal} {J. Phys. A: Math. Theor.}\ }%
  \textbf{\bibinfo {volume} {43}},\ \bibinfo {pages} {165208} (\bibinfo {year}
  {2010})%
  \bibAnnoteFile{NoStop}{hahn-2010}%
\bibitem{beck-2011}%
  \BibitemOpen
  \bibfield{author}{%
  \bibinfo {author} {\bibfnamefont{C.}~\bibnamefont{Beck}},\ }%
  \bibfield{journal}{%
  \bibinfo {journal} {Philos. Trans. Royal Soc. A}\ }%
  \textbf{\bibinfo {volume} {369}},\ \bibinfo {pages} {453} (\bibinfo {year}
  {2011})%
  \bibAnnoteFile{NoStop}{beck-2011}%
\bibitem{beck-2005}%
  \BibitemOpen
  \bibfield{author}{%
  \bibinfo {author} {\bibfnamefont{C.}~\bibnamefont{Beck}}, \bibinfo {author}
  {\bibfnamefont{E.~G.~D.}\ \bibnamefont{Cohen}},\ and\ \bibinfo {author}
  {\bibfnamefont{H.~L.}\ \bibnamefont{Swinney}},\ }%
  \bibfield{journal}{%
  \bibinfo {journal} {Phys. Rev. E}\ }%
  \textbf{\bibinfo {volume} {72}},\ \bibinfo {pages} {056133} (\bibinfo {year}
  {2005})%
  \bibAnnoteFile{NoStop}{beck-2005}%
\bibitem{beck-2005-1}%
  \BibitemOpen
  \bibfield{author}{%
  \bibinfo {author} {\bibfnamefont{C.}~\bibnamefont{Beck}}, \bibinfo {author}
  {\bibfnamefont{E.~G.~D.}\ \bibnamefont{Cohen}},\ and\ \bibinfo {author}
  {\bibfnamefont{S.}~\bibnamefont{Rizzo}},\ }%
  \bibfield{journal}{%
  \bibinfo {journal} {Europhys. News}\ }%
  \textbf{\bibinfo {volume} {36}},\ \bibinfo {pages} {189} (\bibinfo {year}
  {2005})%
  \bibAnnoteFile{NoStop}{beck-2005-1}%
\bibitem{beck-2007}%
  \BibitemOpen
  \bibfield{author}{%
  \bibinfo {author} {\bibfnamefont{C.}~\bibnamefont{Beck}},\ }%
  \bibfield{journal}{%
  \bibinfo {journal} {Phys. Rev. Lett.}\ }%
  \textbf{\bibinfo {volume} {98}},\ \bibinfo {pages} {064502} (\bibinfo {year}
  {2007})%
  \bibAnnoteFile{NoStop}{beck-2007}%
\bibitem{beck-2006}%
  \BibitemOpen
  \bibfield{author}{%
  \bibinfo {author} {\bibfnamefont{C.}~\bibnamefont{Beck}},\ }%
  \bibfield{journal}{%
  \bibinfo {journal} {Physica A}\ }%
  \textbf{\bibinfo {volume} {365}},\ \bibinfo {pages} {96} (\bibinfo {year}
  {2006})%
  \bibAnnoteFile{NoStop}{beck-2006}%
\bibitem{sattin-2002}%
  \BibitemOpen
  \bibfield{author}{%
  \bibinfo {author} {\bibfnamefont{F.}~\bibnamefont{Sattin}}\ and\ \bibinfo
  {author} {\bibfnamefont{L.}~\bibnamefont{Salasnich}},\ }%
  \bibfield{journal}{%
  \bibinfo {journal} {Phys. Rev. E}\ }%
  \textbf{\bibinfo {volume} {65}},\ \bibinfo {pages} {035106(R)} (\bibinfo
  {year} {2002})%
  \bibAnnoteFile{NoStop}{sattin-2002}%
\bibitem{ausloos-2003}%
  \BibitemOpen
  \bibfield{author}{%
  \bibinfo {author} {\bibfnamefont{M.}~\bibnamefont{Ausloos}}\ and\ \bibinfo
  {author} {\bibfnamefont{K.}~\bibnamefont{Ivanova}},\ }%
  \bibfield{journal}{%
  \bibinfo {journal} {Phys. Rev. E}\ }%
  \textbf{\bibinfo {volume} {68}},\ \bibinfo {pages} {046122} (\bibinfo {year}
  {2003})%
  \bibAnnoteFile{NoStop}{ausloos-2003}%
\bibitem{queiros-2005}%
  \BibitemOpen
  \bibfield{author}{%
  \bibinfo {author} {\bibfnamefont{S.~M.~D.}\ \bibnamefont{Queiros}}\ and\
  \bibinfo {author} {\bibfnamefont{C.}~\bibnamefont{Tsallis}},\ }%
  \bibfield{journal}{%
  \bibinfo {journal} {Europhys. Lett.}\ }%
  \textbf{\bibinfo {volume} {69}},\ \bibinfo {pages} {893} (\bibinfo {year}
  {2005})%
  \bibAnnoteFile{NoStop}{queiros-2005}%
\bibitem{queiros-2004}%
  \BibitemOpen
  \bibfield{author}{%
  \bibinfo {author} {\bibfnamefont{S.~M.~D.}\ \bibnamefont{Queiros}},\ }%
  \bibfield{journal}{%
  \bibinfo {journal} {Physica A}\ }%
  \textbf{\bibinfo {volume} {344}},\ \bibinfo {pages} {619} (\bibinfo {year}
  {2004})%
  \bibAnnoteFile{NoStop}{queiros-2004}%
\bibitem{queiros-2005-1}%
  \BibitemOpen
  \bibfield{author}{%
  \bibinfo {author} {\bibfnamefont{S.~M.~D.}\ \bibnamefont{Queiros}}\ and\
  \bibinfo {author} {\bibfnamefont{C.}~\bibnamefont{Tsallis}},\ }%
  \bibfield{journal}{%
  \bibinfo {journal} {Eur. Phys. J. B}\ }%
  \textbf{\bibinfo {volume} {48}},\ \bibinfo {pages} {139} (\bibinfo {year}
  {2005})%
  \bibAnnoteFile{NoStop}{queiros-2005-1}%
\bibitem{queiros-2005-2}%
  \BibitemOpen
  \bibfield{author}{%
  \bibinfo {author} {\bibfnamefont{S.~M.~D.}\ \bibnamefont{Queiros}},\ }%
  \bibfield{journal}{%
  \bibinfo {journal} {Europhys. Lett.}\ }%
  \textbf{\bibinfo {volume} {71}},\ \bibinfo {pages} {339} (\bibinfo {year}
  {2005})%
  \bibAnnoteFile{NoStop}{queiros-2005-2}%
\bibitem{souza-2006}%
  \BibitemOpen
  \bibfield{author}{%
  \bibinfo {author} {\bibfnamefont{J.}~\bibnamefont{de~Souza}}, \bibinfo
  {author} {\bibfnamefont{L.~G.}\ \bibnamefont{Moyano}},\ and\ \bibinfo
  {author} {\bibfnamefont{S.~M.~D.}\ \bibnamefont{Queiros}},\ }%
  \bibfield{journal}{%
  \bibinfo {journal} {Eur. Phys. J. B}\ }%
  \textbf{\bibinfo {volume} {50}},\ \bibinfo {pages} {165} (\bibinfo {year}
  {2006})%
  \bibAnnoteFile{NoStop}{souza-2006}%
\bibitem{ahn-1999}%
  \BibitemOpen
  \bibfield{author}{%
  \bibinfo {author} {\bibfnamefont{D.}~\bibnamefont{Ahn}}\ and\ \bibinfo
  {author} {\bibfnamefont{B.}~\bibnamefont{Gao}},\ }%
  \bibfield{journal}{%
  \bibinfo {journal} {Rev. Fin. Studies}\ }%
  \textbf{\bibinfo {volume} {12}},\ \bibinfo {pages} {721} (\bibinfo {year}
  {1999})%
  \bibAnnoteFile{NoStop}{ahn-1999}%
\bibitem{engle-2001}%
  \BibitemOpen
  \bibfield{author}{%
  \bibinfo {author} {\bibfnamefont{R.~F.}\ \bibnamefont{Engle}}\ and\ \bibinfo
  {author} {\bibfnamefont{A.}~\bibnamefont{Paton}},\ }%
  \bibfield{journal}{%
  \bibinfo {journal} {Quant. Finance}\ }%
  \textbf{\bibinfo {volume} {1}},\ \bibinfo {pages} {237} (\bibinfo {year}
  {2001})%
  \bibAnnoteFile{NoStop}{engle-2001}%
\bibitem{plerou-2001}%
  \BibitemOpen
  \bibfield{author}{%
  \bibinfo {author} {\bibfnamefont{V.}~\bibnamefont{Plerou}}, \bibinfo {author}
  {\bibfnamefont{P.}~\bibnamefont{Gopikrishnan}}, \bibinfo {author}
  {\bibfnamefont{X.}~\bibnamefont{Gabaix}}, \bibinfo {author}
  {\bibfnamefont{L.~A.~N.}\ \bibnamefont{Amaral}},\ and\ \bibinfo {author}
  {\bibfnamefont{H.~E.}\ \bibnamefont{Stanley}},\ }%
  \bibfield{journal}{%
  \bibinfo {journal} {Quant. Finance}\ }%
  \textbf{\bibinfo {volume} {1}},\ \bibinfo {pages} {262} (\bibinfo {year}
  {2001})%
  \bibAnnoteFile{NoStop}{plerou-2001}%
\bibitem{gabaix-2003}%
  \BibitemOpen
  \bibfield{author}{%
  \bibinfo {author} {\bibfnamefont{X.}~\bibnamefont{Gabaix}}, \bibinfo {author}
  {\bibfnamefont{P.}~\bibnamefont{Gopikrishnan}}, \bibinfo {author}
  {\bibfnamefont{V.}~\bibnamefont{Plerou}},\ and\ \bibinfo {author}
  {\bibfnamefont{H.~E.}\ \bibnamefont{Stanley}},\ }%
  \bibfield{journal}{%
  \bibinfo {journal} {Nature}\ }%
  \textbf{\bibinfo {volume} {423}},\ \bibinfo {pages} {267} (\bibinfo {year}
  {2003})%
  \bibAnnoteFile{NoStop}{gabaix-2003}%
\bibitem{Kononovicius2011}%
  \BibitemOpen
  \bibfield{author}{%
  \bibinfo {author} {\bibfnamefont{A.}~\bibnamefont{Kononovi\v{c}ius}}\ and\
  \bibinfo {author} {\bibfnamefont{V.}~\bibnamefont{Gontis}},\ }%
  \bibfield{journal}{%
  \bibinfo {journal} {Physica A}\ }%
  \textbf{\bibinfo {volume} {390}},\ \bibinfo {pages}
  {doi:10.1016/j.physa.2011.08.061} (\bibinfo {year} {2011})%
  \bibAnnoteFile{NoStop}{Kononovicius2011}%
\end{thebibliography}
\end{document}